\documentclass[%
 reprint,
 amsmath,amssymb,
 aps,
superscriptaddress,
]{revtex4-2}

\usepackage{graphicx}
\usepackage{dcolumn}
\usepackage{bm}
\usepackage{color,soul,url}

\begin{document}

\preprint{APS/123-QED}

\title{Quantized Fractional Thouless Pumping of Solitons}

\author{Marius J\"urgensen}
 \email{marius@psu.edu}
 \affiliation{Department of Physics, The Pennsylvania State University, University Park, Pennsylvania 16802, USA}
\author{Sebabrata Mukherjee}
 \affiliation{Department of Physics, The Pennsylvania State University, University Park, Pennsylvania 16802, USA}
\affiliation{Department of Physics, Indian Institute of Science, Bangalore 560012, India}
\author{Christina J\"org}
\author{Mikael C. Rechtsman}%
 \email{mcrworld@psu.edu}
\affiliation{Department of Physics, The Pennsylvania State University, University Park, Pennsylvania 16802, USA}

\date{\today}

\begin{abstract}

In many contexts, the interaction between particles gives rise to emergent and perhaps unanticipated physical phenomena.  An example is the fractional quantum Hall effect, where interaction between electrons gives rise to fractionally quantized Hall conductance. In photonic systems, the nonlinear response of an ambient medium acts to mediate interaction between photons; in the mean-field limit these dynamics are described by the nonlinear Schr\"odinger (also called Gross-Pitaevskii) equation. Recently, it was shown that at weak nonlinearity, soliton motion in nonlinear Thouless pumps (a dimensionally reduced implementation of a Chern insulator) could be quantized to the Chern number of the band from which the soliton bifurcates. Here, we show theoretically and experimentally using arrays of coupled optical waveguides that sufficiently strong nonlinearity acts to fractionally quantize the motion of solitons. Specifically, we find that the soliton returns to itself after multiple cycles of the Thouless pump -- but displaced by an integer number of unit cells -- leading to a rich fractional plateaux structure describing soliton motion. Our results demonstrate a perhaps surprising example of the behavior of non-trivial topological systems in the presence of interactions.

\end{abstract}

\maketitle



\section{Introduction}

In the integer quantum Hall effect, a gas of non-interacting electrons can be rigorously shown to exhibit integer-quantized Hall conductance \cite{Klitzing1980,Thouless1982,Simon1983}, fixed to a topological invariant: the Chern number. It was therefore a surprise when plateaux of fractional conductance appeared in the experiment of Tsui, Stormer and Gossard \cite{Tsui1982}. Here, the interaction between electrons played the key role, giving rise to the formation of fractionally-charged quasiparticles \cite{Laughlin1983anomalous,Jain1989,Jain2007composite}. In addition to electronic systems, topological phenomena have also been predicted and observed in bosonic systems, including with microwave \cite{Wang2009}, and optical photons \cite{Raghu2008,Rechtsman2013,Hafezi2013}, ultracold atoms \cite{Atala2013,Jotzu2014,Aidelsburger2015}, mechanical systems \cite{Suesstrunk2015,Nash2015}, and others. Compared to their electronic counterparts, photonic systems offer rich design flexibility and new physics \cite{Ozawa2019}, such as the inclusion of non-Hermiticity \cite{Zeuner2015}, arbitrary driving, and nonlinearity \cite{Smirnova2020}. However, photons interact extremely weakly in free space, meaning that photon-photon interactions must be mediated by an ambient medium. The first experiments creating photonic two-particle Laughlin states used repulsive interactions mediated by Rydberg atoms in a twisted cavity \cite{Clark2020}. A conceptually different approach has been taken by treating the interactions of many photons in the mean-field limit using nonlinearity. This approach has led to the prediction and observation of various new topological phenomena, such as topological bulk \cite{Lumer2013,Mukherjee2020a} and edge solitons \cite{Ablowitz2014,Leykam2016,Mukherjee2020b}, nonlinearly-induced topological insulators \cite{Maczewsky2020}, and other non-Hermitian and nonlinear phenomena \cite{Xia2020,Xia2021}.

One class of topological models, specifically suitable to study in photonics due to design flexibility in fabricated structures, are Thouless pumps \cite{Thouless1983a,Niu1984}. These models are 1+1 dimensional reductions of Chern insulators, using a time-periodic modulation to emulate a wavevector dimension of the related two-dimensional Chern insulator. Given a uniform band occupation and adiabatic driving, the displacement per period in a Thouless pump is dictated by the Chern number of the occupied band. Thouless pumps have been studied in various systems (see, for example, Refs. \cite{Kraus2012,Lohse2016,Nakajima2016,Lohse2018,Wenchao2018,Fedorova2020,Cerjan2020Thouless,Grinberg2020}). The inclusion of nonlinearity into Thouless pumps has recently led to the discovery of quantized pumping of solitons, where the displacement is dictated by the Chern number of the band from which they bifurcate, despite lacking the notion of a uniformly filled band \cite{Juergensen2021}. Quantization stems from the fact that the soliton comes back to itself -- modulo a translation by an integer number of unit cells due to translation invariance -- after each period. 

Here, we theoretically predict and experimentally demonstrate quantized fractional Thouless pumping of solitons, in which the strength of the nonlinearity exceeds the relevant band gap. We experimentally observe fractional pumping by a fraction of $f$=-1/2 in arrays of evanescently-coupled waveguides with Kerr nonlinearity. Fractional pumping occurs as the wavefunction returns to itself -- modulo a translation by an integer number of unit cells -- only after multiple periods. Finally, we show how tuning the strength of the nonlinearity leads to multiple plateaux of integer and fractional quantized displacement within one Thouless pump model. 
\section{The nonlinear AAH-model}

\begin{figure*}
    \includegraphics{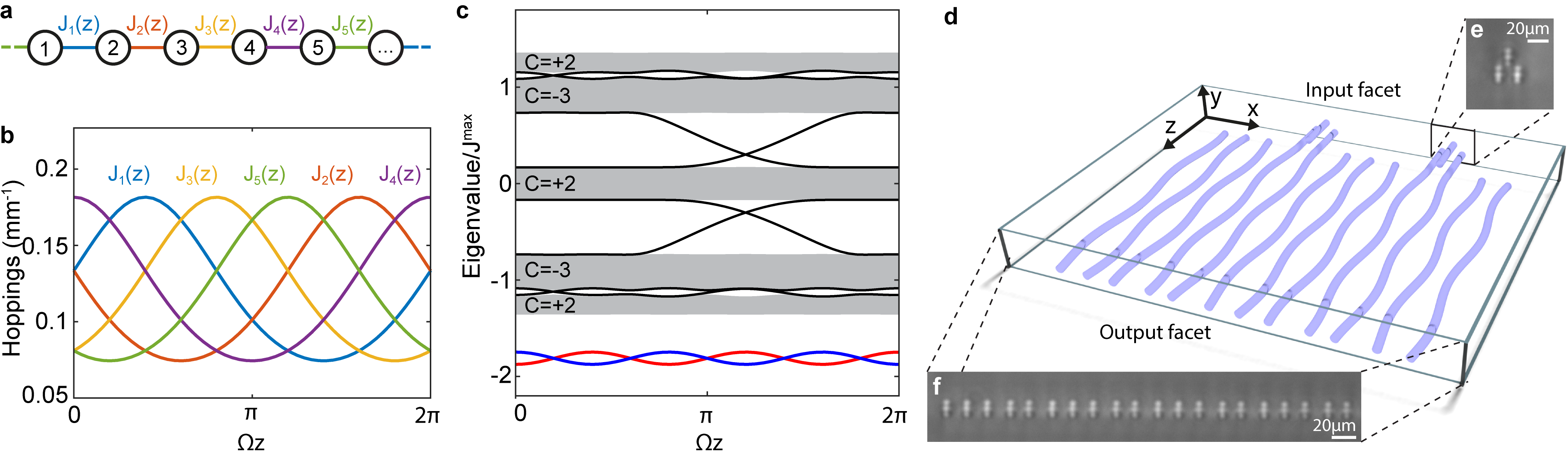}
    \caption{\label{fig1} Model for quantized fractional Thouless pumping. \textbf{a.} Illustration of the off-diagonal AAH-model with five sites per unit cell and $z$-dependent hoppings $J_n(z)$ between nearest-neighbor sites \textbf{b.} Modulation of the hopping strength over one period. \textbf{c.} Band structure (instantaneous energy eigenvalues) of the Hamiltonian showing five bands (grey) with Chern numbers C=\{2,-3,2,-3,2\}, end states (black) crossing the band gaps, and nonlinear eigenvalues (red and blue) of instantaneous solitons that are pumped by a fraction of $f=-1/2$. \textbf{d.} Schematic illustration of the implementation of the model in arrays of evanescently-coupled waveguides. Only two waveguides per unit cell extend to the input facet with an additional waveguide fabricated on top to transform a single-site excitation into an effective two-site excitation. \textbf{e.} and \textbf{f.} White-light micrograph showing the input and output facet, respectively.}
\end{figure*}

Experimentally, we realize a nonlinear Thouless pump by focusing high-peak-power laser pulses into arrays of single-mode evanescently-coupled waveguides. Due to the Kerr effect, the refractive index becomes intensity-dependent and the propagation of photons in the system is described by the discrete nonlinear Schr\"odinger equation \cite{Christodoulides1988, eisenberg1998discrete, fleischer2003observation, christodoulides2003discretizing, Kivshar2003, lederer2008discrete, Kevrekidis2009}
\begin{equation}
i \frac{\partial}{\partial z} \psi_m(z) = \sum_n H_{mn}(z) \psi_n(z) - g
|\psi_m(z)|^2 \psi_m(z)
\end{equation}
where $\psi_n(z)$ is the amplitude of the wavefunction at propagation distance $z$ for site $n$, $H_{mn}(z)$ is a $z$-dependent tight-binding Hamiltonian describing a topological Thouless pump and $g>0$ describes the strength of the focusing Kerr nonlinearity. In waveguide systems, $z$ plays the role of a temporal coordinate. Equation (1) is also known as the Gross-Pitaevskii equation, which describes interacting bosons in a Bose-Einstein condensate in the mean-field limit \cite{Dalfavo1999,Pitaevskii2016Bose}. Therefore, our results are not restricted to photonics but hold for a range of interacting and nonlinear bosonic systems \cite{Donley2001,Abo-Shaeer2001Observation,Deng2002,Kasprzak2006,Balili2007}. In waveguides, the nonlinearity describes an effective interaction between photons mediated by the ambient material, whose strength is given by $g$. To treat experiment and theory on the same footing, we define $P=\sum_n |\psi_n(z=0)|^2$, and refer to the strength of nonlinearity as a dimensionless quantity: $gP/J^{\text{max}}$, where $J^{\text{max}}$ is the largest hopping value in the Hamiltonian (see Supplementary Information Sec. 4 for calibration of $g$).

We illustrate fractional Thouless pumping in an off-diagonal Aubry-Andr\'e-Harper (AAH) model \cite{Harper1955,Aubry1980,Kraus2012,Ke2016} with five sites per unit cell and zero on-site detuning (see Fig. 1a). Its nearest-neighbor couplings $J_n(z)$ are periodically modulated in $z$ and the Hamiltonian is given by $H_{mn}(z) = -J_m(z) \delta_{n,m+1} -J_{m-1} \delta_{n,m-1}$. Fig. 1b shows the strength of the hoppings over one period as used in the experiment. The band structure of this model is depicted in Fig. 1c and has five bands with Chern numbers C=\{2,-3,2,-3,2\}. A schematic illustration of the implementation in an array of evanescently-coupled waveguides is shown in Fig. 1d, where the periodic modulation of the distance between neighboring waveguides changes the evanescent hopping strength. In the experiment the position of waveguide $n$ is $x_n(z) = nd + \delta \cos(\Omega z + \frac{4\pi}{5}n - \frac{6\pi}{20})$, where $d$ defines the average separation between waveguides, $\delta$ is the spatial modulation strength and $\Omega$ is the modulation frequency. Throughout this work we use $d$=17.25\,$\mu$m and $\delta$=1\,$\mu$m. Fig. 1d also schematically shows an input region, where only two waveguides per unit cell are extended all the way to the input facet with an additional waveguide on top (see also Fig. 1e). We use this `triple coupler' to transform a single-site excitation of the upper waveguide into an effective two-site excitation of the two lower waveguides. White light images of the input and output facets are shown in Figs. 1e,f, respectively.

\begin{figure*}
    \includegraphics{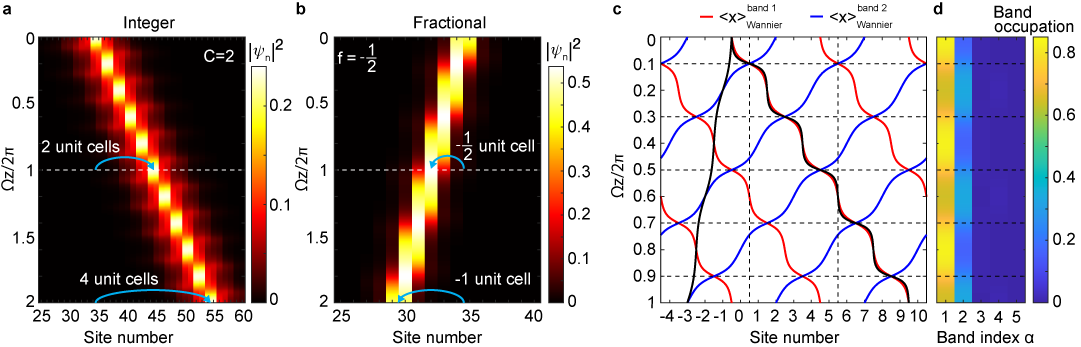}
    \caption{\label{fig2} Theory of quantized fractional Thouless pumping. \textbf{a.} Wavefunction of the instantaneous soliton ($gP/J^{\text{max}}$=0.55) that at low power bifurcates from the lowest band with Chern number $C$=2, calculated for two periods. The displacement per period is two unit cells. \textbf{b.} Similar to \textbf{a}, but for a soliton with $gP/J^{\text{max}}$=1.65. The displacement after one period is one half of a unit cell, and -1 unit cell after two periods. Note that the soliton’s shape has changed after each period. \textbf{c.} Center of mass positions of the Wannier functions of the two lowest bands (red and blue) together with the position of the solitons from \textbf{a} and \textbf{b} (shown in black). While the low-power soliton follows the position of the Wannier function of the lowest band, the path of the fractionally pumped soliton is consistent with the geometric pattern created by the paths of the Wannier functions of the two lowest bands. \textbf{d.} Projected occupation of the linear bands for the fractionally pumped soliton in \textbf{b} over one period, confirming the influence of the second band.}
\end{figure*}

\section{Quantized Fractional Thouless Pumping}

Quantized pumping in linear Thouless pumps ($g=0$) requires a uniform band occupation and adiabatic driving. Under these conditions the displacement per period is dictated by the Chern number of the occupied band.  The scenario is very different in the nonlinear domain, which we explain as follows. Nonlinear systems allow for a spatially localized eigenstate for which the nonlinearity balances the diffraction: a soliton \cite{Askaryan1962,Chiao1964,Ablowitz1981,Christodoulides1988,Stegeman1999}. At low power, solitons can be traced back to the (linear) band from which they bifurcate. Recently, it has been shown that low-power solitons in Thouless pumps move (i.e., are pumped) by the Chern number of the band from which they bifurcate despite nonuniform band occupation \cite{Juergensen2021,Juergensen2021TheChern, Mostaan2021}. Instead, integer quantization occurs because after each period the soliton returns to its initial state but translated by an integer number of unit cells. 

In the adiabatic limit the propagation of a stable soliton can be examined by calculating the instantaneous soliton for each $z$-slice. We focus on the propagation of a soliton that, at $z=0$, bifurcates from the lowest band. Fig. 2a shows the low-power soliton ($gP/J^{\text{max}}$=0.55) for two pumping periods. After each period the soliton’s wavefunction returns to its initial wavefunction, only translated by two unit cells, as dictated by the Chern number of the band from which the soliton bifurcates (the band has Chern number 2). With increasing power, nonlinear eigenstates undergo nonlinear bifurcations, and new propagation paths emerge, as has been illustrated for Thouless pumps in Ref. \cite{Juergensen2021}. The propagation of a soliton for higher power ($gP/J^{\text{max}}$=1.65), which shows fractional pumping, is displayed in Fig. 2b. In this case, the soliton after one period is clearly different from the soliton at $z$=0. It is not peaked on two sites, but instead on a single site and its center of mass displacement is only -1/2 of a unit cell. Only after two periods is the soliton’s wavefunction identical to the initial one modulo a translation by one unit cell in the negative direction (leftward). This behavior goes hand in hand with the presence of two soliton solutions that are degenerate at certain points in the pump cycle (these are the nonlinear eigenvalues plotted in Fig. 1c in red and blue) as we explain below. 

To explain this behavior, we plot in Fig. 2c the center of mass position of the Wannier functions for the two lowest bands (with Chern numbers 2 and -3) together with the positions of the solitons from Figs. 2a and b. As proved in Refs. \cite{Juergensen2021TheChern,Mostaan2021} the stable low-power soliton follows (with small deviations) the position of the Wannier function of the band from which it bifurcates. Hence, the positions of the Wannier functions dictate the paths of the low-power solitons. As the two lowest bands of the AAH-model are only separated by a small band gap, the underlying assumption that the soliton’s dynamics are determined by a single band only is no longer justified for increasing power. Instead, we find that for ($gP/J^{\text{max}}$=1.65), the soliton follows a new path that is compatible with the geometric pattern created by both Wannier centers of the bottom two bands. Importantly, for the new soliton path to occur the previous contiguous path must be broken. Here, nonlinear pitchfork bifurcations act to split the path at positions with high spatial symmetry \cite{Juergensen2021}; those are exactly at the crossing points of the Wannier functions of both bands, where their positions are pinned due to spatial symmetries. The split paths then `recombine' to form a new contiguous path responsible for fractional pumping (see Supplementary Information Sec. 6). Additionally, we verify the influence of the second lowest band by investigating the band occupations (the summed projection coefficients for the five (linear) energy bands), displayed in Fig. 2d. We observe a clear participation of the second-lowest band.

On this basis, we label the fractionally pumped soliton by a fraction $f$: the numerator of the fraction is defined by the number of unit cells by which the soliton is pumped in the $x$-direction before returning to the same wavefunction; the denominator is defined by the number of pump cycles (in $z$) over which this process occurs. For the case shown in Fig. 2b, $f$=(2-3)/2=-1/2, where we have suggestively written the numerator as the summation over the Chern numbers (+2 and -3) of the two participating bands. While this is not generally valid, it applies to our Thouless pump model due to the $z$-symmetric modulation of the hoppings. We point out, that for general models, the fact that $f$=-1/2 does not imply that the displacement after one period is -1/2 unit cells; however, our system shows a displacement of -1/2 unit cells after one cycle due to additional spatial symmetry, in combination with the $z$-symmetric modulation of the coupling functions (Fig. 1b).

\section{Experimental observation of fractional pumping}

\begin{figure*}
    \includegraphics{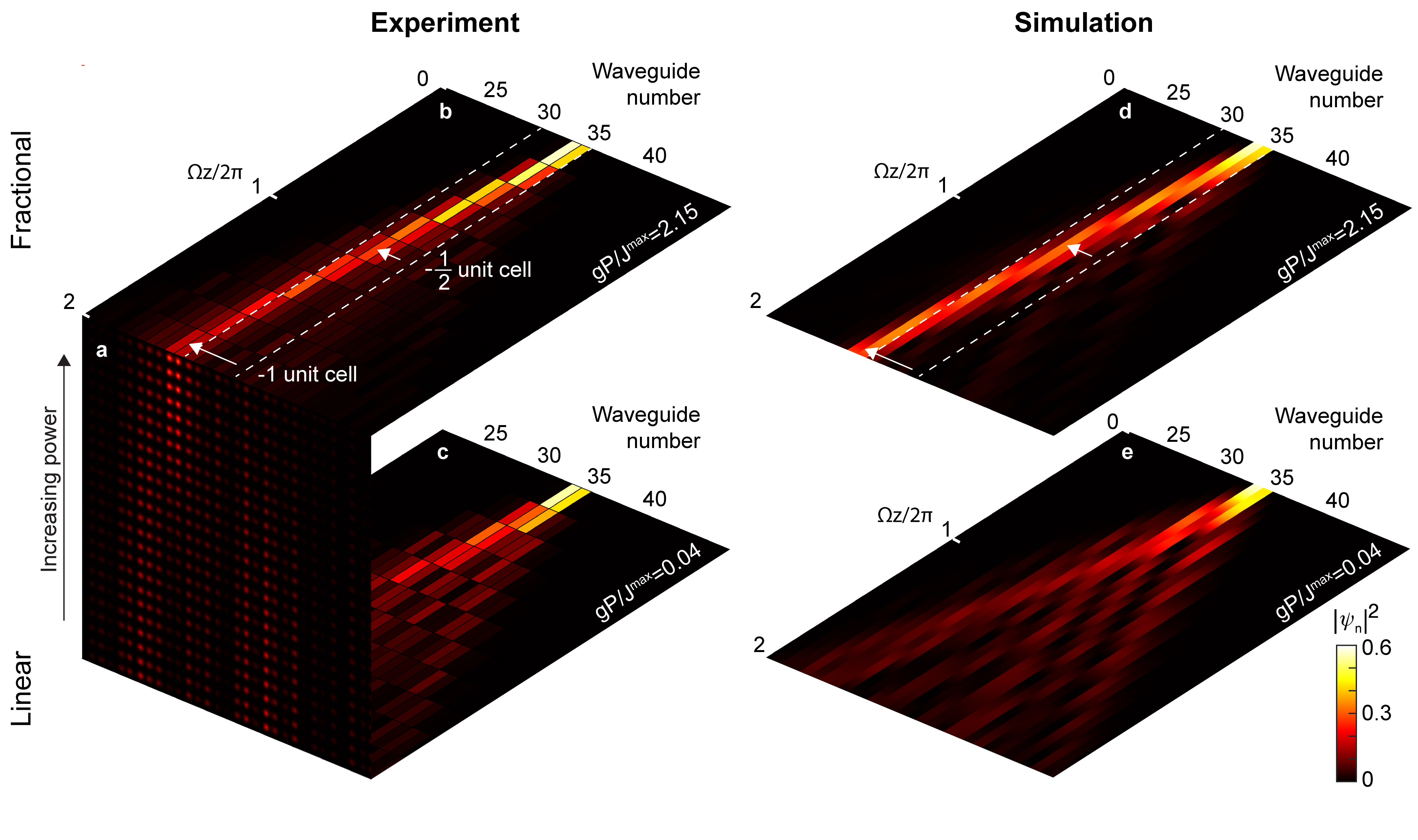}
    \caption{\label{fig3} Experimental observation of a $f$=-1/2 fractionally pumped soliton. \textbf{a.} Output facet showing the intensities in the waveguide modes after propagation for two periods. The bottom row shows the output facet for linear propagation ($gP/J^{\text{max}}$=0.04). The next row was taken for $gP/J^{\text{max}}$=0.09 and then for each row the power was increased in equal steps of $gP/J^{\text{max}} \approx$0.09  until the top row with a maximum input power of $gP/J^{\text{max}}$=2.15. \textbf{b,c.} Normalized integrated intensities in each waveguide for different $z$-slices for an input power of $gP/J^{\text{max}}$=0.04 and $gP/J^{\text{max}}$=2.15 in \textbf{c} and \textbf{b}, respectively. The white dashed lines mark one unit cell. After one period the soliton is peaked on a single site and its center of mass has shifted by half a unit cell. After two periods the soliton is peaked on two sites and displaced by -1 unit cell. \textbf{d,e.} Corresponding tight-binding propagation simulations including losses and using the measured initial two-site excitation. For direct comparison with the experiment the plotted intensities are normalized for each $z$. A comparison of the propagation for all power values is shown in Supplementary Animation 1.}
\end{figure*}

We experimentally observe quantized fractional Thouless pumping in evanescently-coupled waveguide arrays with Kerr nonlinearity. The waveguides are fabricated via femtosecond-direct laser writing in borosilicate glass \cite{Miura1996,Szameit2010}. Straight waveguides show propagation losses of (0.33$\pm$0.02)\,dB/cm and we measure no additional nonlinear losses (see Supplementary Information Sec. 3 and 4). To excite the system, we focus high-power laser pulses into the waveguides, which are temporally stretched to 2\,ps and down-chirped \cite{Mukherjee2020a}. This configuration minimizes the generation of new wavelengths via self-phase modulation while reaching the necessary degree of nonlinearity. Maintaining a narrow spectrum is essential, as the hopping constant is a function of wavelength. In our experiment, the spectrum at the position of the pumped soliton broadens to 14\,nm (see Supplementary Information Sec. 4) for maximum input power and propagation distance. Since the coupling constants between waveguides vary minimally over this range of wavelengths \cite{Mukherjee2020a}, Eq. (1) well-describes our system, provided we include unavoidable propagation losses.

We measure a fractional pumping of $f$=-1/2, meaning that the soliton is pumped to the left by one unit cell after two periods. For our experiment, it is crucial to efficiently excite the soliton, whose wavefunction is mainly peaked on two sites for $z=0$, as shown in Fig. 2b. To facilitate this, our sample contains a 5-mm-long input region, in which only two waveguides per unit cell are extended all the way to the input facet, together with one additional waveguide on top (see also Figs. 1d,e). This `triple coupler' converts a single-site excitation of the upper waveguide into an effective two-site excitation of the lower waveguides (see Supplementary Information Sec. 5).

We detect the fractional pumping behavior of the soliton at high power by measuring the intensity distribution of the waveguide modes at the output facet as a function of the input power for a lattice with 12 unit cells. Figure 3a shows the normalized mode intensities at the output facet after two periods. Each row of modes corresponds to an individual measurement, and the input power increases from bottom to top. For the lowest input power ($gP/J^{\text{max}}$=0.04; bottom row in Fig. 3a), the system behaves linearly, and the intensity diffracts widely in the waveguide array. For increasing input power, the nonlinearity counteracts the diffraction, and less spreading is visible. At the maximum input power of $gP/J^{\text{max}}$=2.15 (top row in Fig. 3a) the intensity at the output facet is localized mainly to two waveguides, one unit cell away from the excited waveguides. This is the signature of the $f=-1/2$ fractionally pumped soliton after two periods.

In order to further verify the fractional pumping behavior, we map out the propagation of the soliton by repeatedly cutting the sample (for further information see Supplementary Information Sec. 2) and imaging the output facet. The normalized integrated intensities per mode (which are equivalent to $|\psi_n|^2$), over two periods are shown in Fig. 3c for the linear propagation ($gP/J^{\text{max}}$=0.04) and in Fig. 3b for $gP/J^{\text{max}}$=2.15. Corresponding numerical simulations using the experimentally measured mode intensities of the effective two-site excitation and including realistic losses are shown in Fig. 3e for the linear and Fig. 3d for the fractionally pumped case. In Supplementary Animation 1 we provide the comparison between theory and experiment for additional input power values; this animation clearly shows the transition from linear diffraction to soliton formation. While here we only present measurements for the fractionally pumped soliton of one unit cell, we found the same behavior in all other measured unit cells.

We point out that our waveguide system does not behave perfectly adiabatically. Therefore, we see radiation from the soliton in both experiment and simulation. In the experiment, the contrast is further lowered by the tails of the laser pulses, which have lower intensity and thus behave more linearly. Nonetheless, our experiment clearly shows a soliton displaced to the left by one unit cell after two periods, with the characteristic shape of being localized on two sites after two periods and localized on just one site after one period (see also the instantaneous soliton in Fig. 2b).

\section{Multiple fractional plateaux}

\begin{figure}
    \includegraphics{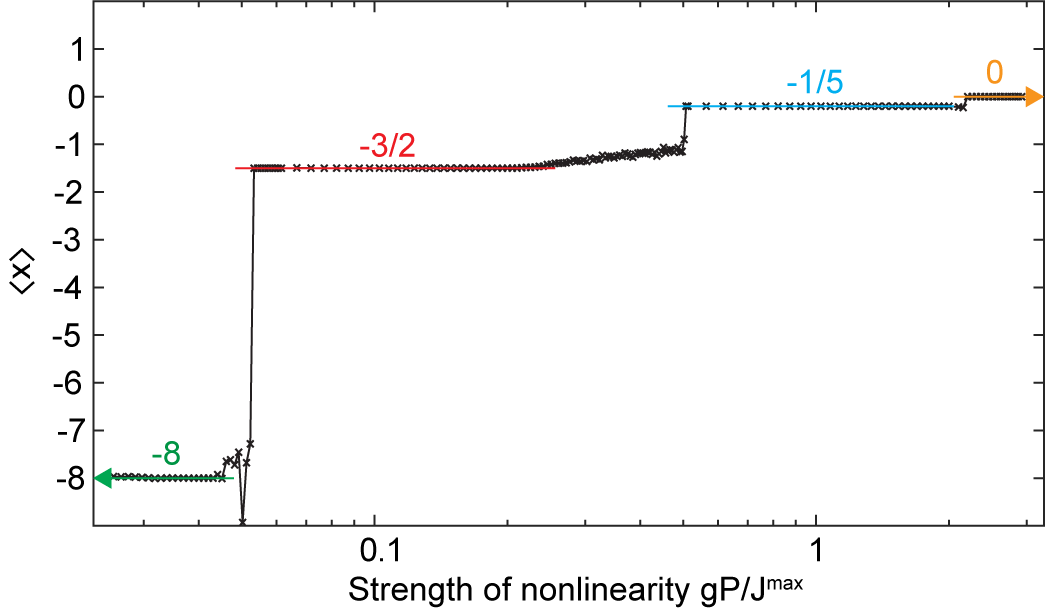}
    \caption{\label{fig4} Average center of mass displacement per period, $\langle x \rangle$, calculated for an off-diagonal AAH-model with 13 bands. For low power ($gP/J^{\text{max}}<$0.04) the soliton motion is determined by the integer quantized Chern number of the band from which the soliton bifurcates ($C=-8$). With increasing nonlinearity plateaux of fractionally quantized displacement of -3/2 and -1/5 appear. For $gP/J^{\text{max}}>$2.2, the soliton is trapped and the average displacement is zero. Note, that data points for $gP/J^{\text{max}}<$0.06 and $gP/J^{\text{max}}>$0.06 are calculated differently to ensure adiabaticity (see Supplementary Information Sec. 7).}
\end{figure}

Finally, we demonstrate numerically that multiple plateaux of integer and fractionally quantized pumping can occur within one Thouless pump model for increasing nonlinearity. We numerically solve Eq. (1) for an off-diagonal AAH-model with 13 sites per unit cell (see Supplementary Information Sec. 7 for more details on the model and propagation parameters) using periodic boundary conditions. The initial excitation is chosen to be a power-dependent, instantaneous soliton that bifurcates from the lowest band. In Fig. 4 we plot the average center of mass displacement per period $\langle x \rangle$ showing four plateaux of quantized displacement. At low power the center of mass displacement is integer-quantized as dictated by the Chern number of the band from which the soliton bifurcates: $C$=-8. With increasing nonlinearity, the two lowest bands participate in the soliton pumping, resulting in a fractional pumping of $f$=(-8+5)/2=-3/2. At even higher power another fractionally pumped soliton emerges, which -- for this model with $z$-symmetric hopping modulation -- is accurately described by the participation of the five lowest bands: $f$=(-8+5+5-8+5)/5=-1/5. Finally, at very high power the soliton is trapped as the strong nonlinearly-induced on-site detuning effectively detaches the soliton from the lattice. Therefore, the soliton’s displacement per period is zero \cite{Juergensen2021}. In our framework this corresponds to the Chern number average of all bands, which is known to be zero in tight-binding models.

\section{Summary \& Outlook}
In summary, we have theoretically predicted fractional pumping of solitons in nonlinear photonic Thouless pumps and experimentally observed a fractional $f$=-1/2 displacement. Furthermore, we have numerically shown the occurrence of a rich structure of multiple quantized plateaux of integer and fractional displacement. This is evocative of the fractional quantum Hall effect for electrons; the connection is particularly compelling in that both phenomena seem to require degenerate ground states (solitons residing below the lowest band are nonlinear ground states). That said, our results here occur for effectively attractive bosons described by a localized mean-field single-particle wavefunction. This result implies that the fractionalization of transport in interacting topological systems is perhaps more general than was previously understood in the context of the fractional quantum Hall effect.


\vspace{15px}
While finalizing the experiments, the authors became aware of a related work. \cite{Fu2021}

\begin{acknowledgments}
We acknowledge fruitful discussions with S. Gopalakrishnan. We further acknowledge the support of the ONR YIP program under award number N00014-18-1-2595, the ONR-MURI program N00014-20-1-2325, as well as by the Packard Foundation, fellowship number 2017-66821. C.J. gratefully acknowledges funding from the Alexander von Humboldt Foundation within the Feodor-Lynen Fellowship program. Numerical calculations were performed on the Pennsylvania State University’s Institute for Computational and Data Sciences’ Roar supercomputer.
\end{acknowledgments}


\bibliography{bibliography}

\begin{thebibliography}{65}%
\makeatletter
\providecommand \@ifxundefined [1]{%
 \@ifx{#1\undefined}
}%
\providecommand \@ifnum [1]{%
 \ifnum #1\expandafter \@firstoftwo
 \else \expandafter \@secondoftwo
 \fi
}%
\providecommand \@ifx [1]{%
 \ifx #1\expandafter \@firstoftwo
 \else \expandafter \@secondoftwo
 \fi
}%
\providecommand \natexlab [1]{#1}%
\providecommand \enquote  [1]{``#1''}%
\providecommand \bibnamefont  [1]{#1}%
\providecommand \bibfnamefont [1]{#1}%
\providecommand \citenamefont [1]{#1}%
\providecommand \href@noop [0]{\@secondoftwo}%
\providecommand \href [0]{\begingroup \@sanitize@url \@href}%
\providecommand \@href[1]{\@@startlink{#1}\@@href}%
\providecommand \@@href[1]{\endgroup#1\@@endlink}%
\providecommand \@sanitize@url [0]{\catcode `\\12\catcode `\$12\catcode
  `\&12\catcode `\#12\catcode `\^12\catcode `\_12\catcode `\%12\relax}%
\providecommand \@@startlink[1]{}%
\providecommand \@@endlink[0]{}%
\providecommand \url  [0]{\begingroup\@sanitize@url \@url }%
\providecommand \@url [1]{\endgroup\@href {#1}{\urlprefix }}%
\providecommand \urlprefix  [0]{URL }%
\providecommand \Eprint [0]{\href }%
\providecommand \doibase [0]{https://doi.org/}%
\providecommand \selectlanguage [0]{\@gobble}%
\providecommand \bibinfo  [0]{\@secondoftwo}%
\providecommand \bibfield  [0]{\@secondoftwo}%
\providecommand \translation [1]{[#1]}%
\providecommand \BibitemOpen [0]{}%
\providecommand \bibitemStop [0]{}%
\providecommand \bibitemNoStop [0]{.\EOS\space}%
\providecommand \EOS [0]{\spacefactor3000\relax}%
\providecommand \BibitemShut  [1]{\csname bibitem#1\endcsname}%
\let\auto@bib@innerbib\@empty
\bibitem [{\citenamefont {Klitzing}\ \emph {et~al.}(1980)\citenamefont
  {Klitzing}, \citenamefont {Dorda},\ and\ \citenamefont
  {Pepper}}]{Klitzing1980}%
  \BibitemOpen
  \bibfield  {author} {\bibinfo {author} {\bibfnamefont {K.~v.}\ \bibnamefont
  {Klitzing}}, \bibinfo {author} {\bibfnamefont {G.}~\bibnamefont {Dorda}},\
  and\ \bibinfo {author} {\bibfnamefont {M.}~\bibnamefont {Pepper}},\
  }\bibfield  {title} {\bibinfo {title} {New method for high-accuracy
  determination of the fine-structure constant based on quantized hall
  resistance},\ }\href {https://doi.org/10.1103/PhysRevLett.45.494} {\bibfield
  {journal} {\bibinfo  {journal} {Physical Review Letters}\ }\textbf {\bibinfo
  {volume} {45}},\ \bibinfo {pages} {494} (\bibinfo {year} {1980})}\BibitemShut
  {NoStop}%
\bibitem [{\citenamefont {Thouless}\ \emph {et~al.}(1982)\citenamefont
  {Thouless}, \citenamefont {Kohmoto}, \citenamefont {Nightingale},\ and\
  \citenamefont {den Nijs}}]{Thouless1982}%
  \BibitemOpen
  \bibfield  {author} {\bibinfo {author} {\bibfnamefont {D.~J.}\ \bibnamefont
  {Thouless}}, \bibinfo {author} {\bibfnamefont {M.}~\bibnamefont {Kohmoto}},
  \bibinfo {author} {\bibfnamefont {M.~P.}\ \bibnamefont {Nightingale}},\ and\
  \bibinfo {author} {\bibfnamefont {M.}~\bibnamefont {den Nijs}},\ }\bibfield
  {title} {\bibinfo {title} {Quantized hall conductance in a two-dimensional
  periodic potential},\ }\href@noop {} {\bibfield  {journal} {\bibinfo
  {journal} {Physical review letters}\ }\textbf {\bibinfo {volume} {49}},\
  \bibinfo {pages} {405} (\bibinfo {year} {1982})}\BibitemShut {NoStop}%
\bibitem [{\citenamefont {Simon}(1983)}]{Simon1983}%
  \BibitemOpen
  \bibfield  {author} {\bibinfo {author} {\bibfnamefont {B.}~\bibnamefont
  {Simon}},\ }\bibfield  {title} {\bibinfo {title} {Holonomy, the quantum
  adiabatic theorem, and berry's phase},\ }\href
  {https://doi.org/10.1103/PhysRevLett.51.2167} {\bibfield  {journal} {\bibinfo
   {journal} {Physical Review Letters}\ }\textbf {\bibinfo {volume} {51}},\
  \bibinfo {pages} {2167} (\bibinfo {year} {1983})}\BibitemShut {NoStop}%
\bibitem [{\citenamefont {Tsui}\ \emph {et~al.}(1982)\citenamefont {Tsui},
  \citenamefont {Stormer},\ and\ \citenamefont {Gossard}}]{Tsui1982}%
  \BibitemOpen
  \bibfield  {author} {\bibinfo {author} {\bibfnamefont {D.~C.}\ \bibnamefont
  {Tsui}}, \bibinfo {author} {\bibfnamefont {H.~L.}\ \bibnamefont {Stormer}},\
  and\ \bibinfo {author} {\bibfnamefont {A.~C.}\ \bibnamefont {Gossard}},\
  }\bibfield  {title} {\bibinfo {title} {Two-dimensional magnetotransport in
  the extreme quantum limit},\ }\href
  {https://doi.org/10.1103/PhysRevLett.48.1559} {\bibfield  {journal} {\bibinfo
   {journal} {Physical Review Letters}\ }\textbf {\bibinfo {volume} {48}},\
  \bibinfo {pages} {1559} (\bibinfo {year} {1982})}\BibitemShut {NoStop}%
\bibitem [{\citenamefont {Laughlin}(1983)}]{Laughlin1983anomalous}%
  \BibitemOpen
  \bibfield  {author} {\bibinfo {author} {\bibfnamefont {R.~B.}\ \bibnamefont
  {Laughlin}},\ }\bibfield  {title} {\bibinfo {title} {Anomalous quantum hall
  effect: An incompressible quantum fluid with fractionally charged
  excitations},\ }\href {https://doi.org/10.1103/PhysRevLett.50.1395}
  {\bibfield  {journal} {\bibinfo  {journal} {Physical Review Letters}\
  }\textbf {\bibinfo {volume} {50}},\ \bibinfo {pages} {1395} (\bibinfo {year}
  {1983})}\BibitemShut {NoStop}%
\bibitem [{\citenamefont {Jain}(1989)}]{Jain1989}%
  \BibitemOpen
  \bibfield  {author} {\bibinfo {author} {\bibfnamefont {J.~K.}\ \bibnamefont
  {Jain}},\ }\bibfield  {title} {\bibinfo {title} {Composite-fermion approach
  for the fractional quantum hall effect},\ }\href
  {https://doi.org/10.1103/PhysRevLett.63.199} {\bibfield  {journal} {\bibinfo
  {journal} {Physical Review Letters}\ }\textbf {\bibinfo {volume} {63}},\
  \bibinfo {pages} {199} (\bibinfo {year} {1989})}\BibitemShut {NoStop}%
\bibitem [{\citenamefont {Jain}(2007)}]{Jain2007composite}%
  \BibitemOpen
  \bibfield  {author} {\bibinfo {author} {\bibfnamefont {J.~K.}\ \bibnamefont
  {Jain}},\ }\href@noop {} {\emph {\bibinfo {title} {Composite fermions}}}\
  (\bibinfo  {publisher} {Cambridge University Press},\ \bibinfo {year}
  {2007})\BibitemShut {NoStop}%
\bibitem [{\citenamefont {Wang}\ \emph {et~al.}(2009)\citenamefont {Wang},
  \citenamefont {Chong}, \citenamefont {Joannopoulos},\ and\ \citenamefont
  {Soljačić}}]{Wang2009}%
  \BibitemOpen
  \bibfield  {author} {\bibinfo {author} {\bibfnamefont {Z.}~\bibnamefont
  {Wang}}, \bibinfo {author} {\bibfnamefont {Y.}~\bibnamefont {Chong}},
  \bibinfo {author} {\bibfnamefont {J.~D.}\ \bibnamefont {Joannopoulos}},\ and\
  \bibinfo {author} {\bibfnamefont {M.}~\bibnamefont {Soljačić}},\ }\bibfield
   {title} {\bibinfo {title} {Observation of unidirectional
  backscattering-immune topological electromagnetic states},\ }\href
  {https://doi.org/10.1038/nature08293} {\bibfield  {journal} {\bibinfo
  {journal} {Nature}\ }\textbf {\bibinfo {volume} {461}},\ \bibinfo {pages}
  {772} (\bibinfo {year} {2009})}\BibitemShut {NoStop}%
\bibitem [{\citenamefont {Raghu}\ and\ \citenamefont
  {Haldane}(2008)}]{Raghu2008}%
  \BibitemOpen
  \bibfield  {author} {\bibinfo {author} {\bibfnamefont {S.}~\bibnamefont
  {Raghu}}\ and\ \bibinfo {author} {\bibfnamefont {F.~D.~M.}\ \bibnamefont
  {Haldane}},\ }\bibfield  {title} {\bibinfo {title} {Analogs of
  quantum-hall-effect edge states in photonic crystals},\ }\href
  {https://doi.org/10.1103/PhysRevA.78.033834} {\bibfield  {journal} {\bibinfo
  {journal} {Physical Review A}\ }\textbf {\bibinfo {volume} {78}},\ \bibinfo
  {pages} {033834} (\bibinfo {year} {2008})}\BibitemShut {NoStop}%
\bibitem [{\citenamefont {Rechtsman}\ \emph {et~al.}(2013)\citenamefont
  {Rechtsman}, \citenamefont {Zeuner}, \citenamefont {Plotnik}, \citenamefont
  {Lumer}, \citenamefont {Podolsky}, \citenamefont {Dreisow}, \citenamefont
  {Nolte}, \citenamefont {Segev},\ and\ \citenamefont
  {Szameit}}]{Rechtsman2013}%
  \BibitemOpen
  \bibfield  {author} {\bibinfo {author} {\bibfnamefont {M.~C.}\ \bibnamefont
  {Rechtsman}}, \bibinfo {author} {\bibfnamefont {J.~M.}\ \bibnamefont
  {Zeuner}}, \bibinfo {author} {\bibfnamefont {Y.}~\bibnamefont {Plotnik}},
  \bibinfo {author} {\bibfnamefont {Y.}~\bibnamefont {Lumer}}, \bibinfo
  {author} {\bibfnamefont {D.}~\bibnamefont {Podolsky}}, \bibinfo {author}
  {\bibfnamefont {F.}~\bibnamefont {Dreisow}}, \bibinfo {author} {\bibfnamefont
  {S.}~\bibnamefont {Nolte}}, \bibinfo {author} {\bibfnamefont
  {M.}~\bibnamefont {Segev}},\ and\ \bibinfo {author} {\bibfnamefont
  {A.}~\bibnamefont {Szameit}},\ }\bibfield  {title} {\bibinfo {title}
  {Photonic floquet topological insulators},\ }\href
  {https://doi.org/10.1038/nature12066
  https://www.nature.com/articles/nature12066#supplementary-information}
  {\bibfield  {journal} {\bibinfo  {journal} {Nature}\ }\textbf {\bibinfo
  {volume} {496}},\ \bibinfo {pages} {196} (\bibinfo {year}
  {2013})}\BibitemShut {NoStop}%
\bibitem [{\citenamefont {Hafezi}\ \emph {et~al.}(2013)\citenamefont {Hafezi},
  \citenamefont {Mittal}, \citenamefont {Fan}, \citenamefont {Migdall},\ and\
  \citenamefont {Taylor}}]{Hafezi2013}%
  \BibitemOpen
  \bibfield  {author} {\bibinfo {author} {\bibfnamefont {M.}~\bibnamefont
  {Hafezi}}, \bibinfo {author} {\bibfnamefont {S.}~\bibnamefont {Mittal}},
  \bibinfo {author} {\bibfnamefont {J.}~\bibnamefont {Fan}}, \bibinfo {author}
  {\bibfnamefont {A.}~\bibnamefont {Migdall}},\ and\ \bibinfo {author}
  {\bibfnamefont {J.~M.}\ \bibnamefont {Taylor}},\ }\bibfield  {title}
  {\bibinfo {title} {Imaging topological edge states in silicon photonics},\
  }\href {https://doi.org/10.1038/nphoton.2013.274
  https://www.nature.com/articles/nphoton.2013.274#supplementary-information}
  {\bibfield  {journal} {\bibinfo  {journal} {Nature Photonics}\ }\textbf
  {\bibinfo {volume} {7}},\ \bibinfo {pages} {1001} (\bibinfo {year}
  {2013})}\BibitemShut {NoStop}%
\bibitem [{\citenamefont {Atala}\ \emph {et~al.}(2013)\citenamefont {Atala},
  \citenamefont {Aidelsburger}, \citenamefont {Barreiro}, \citenamefont
  {Abanin}, \citenamefont {Kitagawa}, \citenamefont {Demler},\ and\
  \citenamefont {Bloch}}]{Atala2013}%
  \BibitemOpen
  \bibfield  {author} {\bibinfo {author} {\bibfnamefont {M.}~\bibnamefont
  {Atala}}, \bibinfo {author} {\bibfnamefont {M.}~\bibnamefont {Aidelsburger}},
  \bibinfo {author} {\bibfnamefont {J.~T.}\ \bibnamefont {Barreiro}}, \bibinfo
  {author} {\bibfnamefont {D.}~\bibnamefont {Abanin}}, \bibinfo {author}
  {\bibfnamefont {T.}~\bibnamefont {Kitagawa}}, \bibinfo {author}
  {\bibfnamefont {E.}~\bibnamefont {Demler}},\ and\ \bibinfo {author}
  {\bibfnamefont {I.}~\bibnamefont {Bloch}},\ }\bibfield  {title} {\bibinfo
  {title} {Direct measurement of the zak phase in topological bloch bands},\
  }\href {https://doi.org/10.1038/nphys2790
  https://www.nature.com/articles/nphys2790#supplementary-information}
  {\bibfield  {journal} {\bibinfo  {journal} {Nature Physics}\ }\textbf
  {\bibinfo {volume} {9}},\ \bibinfo {pages} {795} (\bibinfo {year}
  {2013})}\BibitemShut {NoStop}%
\bibitem [{\citenamefont {Jotzu}\ \emph {et~al.}(2014)\citenamefont {Jotzu},
  \citenamefont {Messer}, \citenamefont {Desbuquois}, \citenamefont {Lebrat},
  \citenamefont {Uehlinger}, \citenamefont {Greif},\ and\ \citenamefont
  {Esslinger}}]{Jotzu2014}%
  \BibitemOpen
  \bibfield  {author} {\bibinfo {author} {\bibfnamefont {G.}~\bibnamefont
  {Jotzu}}, \bibinfo {author} {\bibfnamefont {M.}~\bibnamefont {Messer}},
  \bibinfo {author} {\bibfnamefont {R.}~\bibnamefont {Desbuquois}}, \bibinfo
  {author} {\bibfnamefont {M.}~\bibnamefont {Lebrat}}, \bibinfo {author}
  {\bibfnamefont {T.}~\bibnamefont {Uehlinger}}, \bibinfo {author}
  {\bibfnamefont {D.}~\bibnamefont {Greif}},\ and\ \bibinfo {author}
  {\bibfnamefont {T.}~\bibnamefont {Esslinger}},\ }\bibfield  {title} {\bibinfo
  {title} {Experimental realization of the topological haldane model with
  ultracold fermions},\ }\href {https://doi.org/10.1038/nature13915} {\bibfield
   {journal} {\bibinfo  {journal} {Nature}\ }\textbf {\bibinfo {volume}
  {515}},\ \bibinfo {pages} {237} (\bibinfo {year} {2014})}\BibitemShut
  {NoStop}%
\bibitem [{\citenamefont {Aidelsburger}\ \emph {et~al.}(2015)\citenamefont
  {Aidelsburger}, \citenamefont {Lohse}, \citenamefont {Schweizer},
  \citenamefont {Atala}, \citenamefont {Barreiro}, \citenamefont {Nascimbène},
  \citenamefont {Cooper}, \citenamefont {Bloch},\ and\ \citenamefont
  {Goldman}}]{Aidelsburger2015}%
  \BibitemOpen
  \bibfield  {author} {\bibinfo {author} {\bibfnamefont {M.}~\bibnamefont
  {Aidelsburger}}, \bibinfo {author} {\bibfnamefont {M.}~\bibnamefont {Lohse}},
  \bibinfo {author} {\bibfnamefont {C.}~\bibnamefont {Schweizer}}, \bibinfo
  {author} {\bibfnamefont {M.}~\bibnamefont {Atala}}, \bibinfo {author}
  {\bibfnamefont {J.~T.}\ \bibnamefont {Barreiro}}, \bibinfo {author}
  {\bibfnamefont {S.}~\bibnamefont {Nascimbène}}, \bibinfo {author}
  {\bibfnamefont {N.~R.}\ \bibnamefont {Cooper}}, \bibinfo {author}
  {\bibfnamefont {I.}~\bibnamefont {Bloch}},\ and\ \bibinfo {author}
  {\bibfnamefont {N.}~\bibnamefont {Goldman}},\ }\bibfield  {title} {\bibinfo
  {title} {Measuring the chern number of hofstadter bands with ultracold
  bosonic atoms},\ }\href {https://doi.org/10.1038/nphys3171} {\bibfield
  {journal} {\bibinfo  {journal} {Nature Physics}\ }\textbf {\bibinfo {volume}
  {11}},\ \bibinfo {pages} {162} (\bibinfo {year} {2015})}\BibitemShut
  {NoStop}%
\bibitem [{\citenamefont {Süsstrunk}\ and\ \citenamefont
  {Huber}(2015)}]{Suesstrunk2015}%
  \BibitemOpen
  \bibfield  {author} {\bibinfo {author} {\bibfnamefont {R.}~\bibnamefont
  {Süsstrunk}}\ and\ \bibinfo {author} {\bibfnamefont {S.~D.}\ \bibnamefont
  {Huber}},\ }\bibfield  {title} {\bibinfo {title} {Observation of phononic
  helical edge states in a mechanical topological insulator},\ }\href
  {https://doi.org/doi:10.1126/science.aab0239} {\bibfield  {journal} {\bibinfo
   {journal} {Science}\ }\textbf {\bibinfo {volume} {349}},\ \bibinfo {pages}
  {47} (\bibinfo {year} {2015})}\BibitemShut {NoStop}%
\bibitem [{\citenamefont {Nash}\ \emph {et~al.}(2015)\citenamefont {Nash},
  \citenamefont {Kleckner}, \citenamefont {Read}, \citenamefont {Vitelli},
  \citenamefont {Turner},\ and\ \citenamefont {Irvine}}]{Nash2015}%
  \BibitemOpen
  \bibfield  {author} {\bibinfo {author} {\bibfnamefont {L.~M.}\ \bibnamefont
  {Nash}}, \bibinfo {author} {\bibfnamefont {D.}~\bibnamefont {Kleckner}},
  \bibinfo {author} {\bibfnamefont {A.}~\bibnamefont {Read}}, \bibinfo {author}
  {\bibfnamefont {V.}~\bibnamefont {Vitelli}}, \bibinfo {author} {\bibfnamefont
  {A.~M.}\ \bibnamefont {Turner}},\ and\ \bibinfo {author} {\bibfnamefont
  {W.~T.~M.}\ \bibnamefont {Irvine}},\ }\bibfield  {title} {\bibinfo {title}
  {Topological mechanics of gyroscopic metamaterials},\ }\href
  {https://doi.org/10.1073/pnas.1507413112} {\bibfield  {journal} {\bibinfo
  {journal} {Proceedings of the National Academy of Sciences}\ }\textbf
  {\bibinfo {volume} {112}},\ \bibinfo {pages} {14495} (\bibinfo {year}
  {2015})}\BibitemShut {NoStop}%
\bibitem [{\citenamefont {Ozawa}\ \emph {et~al.}(2019)\citenamefont {Ozawa},
  \citenamefont {Price}, \citenamefont {Amo}, \citenamefont {Goldman},
  \citenamefont {Hafezi}, \citenamefont {Lu}, \citenamefont {Rechtsman},
  \citenamefont {Schuster}, \citenamefont {Simon}, \citenamefont {Zilberberg},\
  and\ \citenamefont {Carusotto}}]{Ozawa2019}%
  \BibitemOpen
  \bibfield  {author} {\bibinfo {author} {\bibfnamefont {T.}~\bibnamefont
  {Ozawa}}, \bibinfo {author} {\bibfnamefont {H.~M.}\ \bibnamefont {Price}},
  \bibinfo {author} {\bibfnamefont {A.}~\bibnamefont {Amo}}, \bibinfo {author}
  {\bibfnamefont {N.}~\bibnamefont {Goldman}}, \bibinfo {author} {\bibfnamefont
  {M.}~\bibnamefont {Hafezi}}, \bibinfo {author} {\bibfnamefont
  {L.}~\bibnamefont {Lu}}, \bibinfo {author} {\bibfnamefont {M.~C.}\
  \bibnamefont {Rechtsman}}, \bibinfo {author} {\bibfnamefont {D.}~\bibnamefont
  {Schuster}}, \bibinfo {author} {\bibfnamefont {J.}~\bibnamefont {Simon}},
  \bibinfo {author} {\bibfnamefont {O.}~\bibnamefont {Zilberberg}},\ and\
  \bibinfo {author} {\bibfnamefont {I.}~\bibnamefont {Carusotto}},\ }\bibfield
  {title} {\bibinfo {title} {Topological photonics},\ }\href
  {https://doi.org/10.1103/RevModPhys.91.015006} {\bibfield  {journal}
  {\bibinfo  {journal} {Reviews of Modern Physics}\ }\textbf {\bibinfo {volume}
  {91}},\ \bibinfo {pages} {015006} (\bibinfo {year} {2019})}\BibitemShut
  {NoStop}%
\bibitem [{\citenamefont {Zeuner}\ \emph {et~al.}(2015)\citenamefont {Zeuner},
  \citenamefont {Rechtsman}, \citenamefont {Plotnik}, \citenamefont {Lumer},
  \citenamefont {Nolte}, \citenamefont {Rudner}, \citenamefont {Segev},\ and\
  \citenamefont {Szameit}}]{Zeuner2015}%
  \BibitemOpen
  \bibfield  {author} {\bibinfo {author} {\bibfnamefont {J.~M.}\ \bibnamefont
  {Zeuner}}, \bibinfo {author} {\bibfnamefont {M.~C.}\ \bibnamefont
  {Rechtsman}}, \bibinfo {author} {\bibfnamefont {Y.}~\bibnamefont {Plotnik}},
  \bibinfo {author} {\bibfnamefont {Y.}~\bibnamefont {Lumer}}, \bibinfo
  {author} {\bibfnamefont {S.}~\bibnamefont {Nolte}}, \bibinfo {author}
  {\bibfnamefont {M.~S.}\ \bibnamefont {Rudner}}, \bibinfo {author}
  {\bibfnamefont {M.}~\bibnamefont {Segev}},\ and\ \bibinfo {author}
  {\bibfnamefont {A.}~\bibnamefont {Szameit}},\ }\bibfield  {title} {\bibinfo
  {title} {Observation of a topological transition in the bulk of a
  non-hermitian system},\ }\href
  {https://doi.org/10.1103/PhysRevLett.115.040402} {\bibfield  {journal}
  {\bibinfo  {journal} {Physical Review Letters}\ }\textbf {\bibinfo {volume}
  {115}},\ \bibinfo {pages} {040402} (\bibinfo {year} {2015})}\BibitemShut
  {NoStop}%
\bibitem [{\citenamefont {Smirnova}\ \emph {et~al.}(2020)\citenamefont
  {Smirnova}, \citenamefont {Leykam}, \citenamefont {Chong},\ and\
  \citenamefont {Kivshar}}]{Smirnova2020}%
  \BibitemOpen
  \bibfield  {author} {\bibinfo {author} {\bibfnamefont {D.}~\bibnamefont
  {Smirnova}}, \bibinfo {author} {\bibfnamefont {D.}~\bibnamefont {Leykam}},
  \bibinfo {author} {\bibfnamefont {Y.}~\bibnamefont {Chong}},\ and\ \bibinfo
  {author} {\bibfnamefont {Y.}~\bibnamefont {Kivshar}},\ }\bibfield  {title}
  {\bibinfo {title} {Nonlinear topological photonics},\ }\href
  {https://doi.org/10.1063/1.5142397} {\bibfield  {journal} {\bibinfo
  {journal} {Applied Physics Reviews}\ }\textbf {\bibinfo {volume} {7}},\
  \bibinfo {pages} {021306} (\bibinfo {year} {2020})}\BibitemShut {NoStop}%
\bibitem [{\citenamefont {Clark}\ \emph {et~al.}(2020)\citenamefont {Clark},
  \citenamefont {Schine}, \citenamefont {Baum}, \citenamefont {Jia},\ and\
  \citenamefont {Simon}}]{Clark2020}%
  \BibitemOpen
  \bibfield  {author} {\bibinfo {author} {\bibfnamefont {L.~W.}\ \bibnamefont
  {Clark}}, \bibinfo {author} {\bibfnamefont {N.}~\bibnamefont {Schine}},
  \bibinfo {author} {\bibfnamefont {C.}~\bibnamefont {Baum}}, \bibinfo {author}
  {\bibfnamefont {N.}~\bibnamefont {Jia}},\ and\ \bibinfo {author}
  {\bibfnamefont {J.}~\bibnamefont {Simon}},\ }\bibfield  {title} {\bibinfo
  {title} {Observation of laughlin states made of light},\ }\href
  {https://doi.org/10.1038/s41586-020-2318-5} {\bibfield  {journal} {\bibinfo
  {journal} {Nature}\ }\textbf {\bibinfo {volume} {582}},\ \bibinfo {pages}
  {41} (\bibinfo {year} {2020})}\BibitemShut {NoStop}%
\bibitem [{\citenamefont {Lumer}\ \emph {et~al.}(2013)\citenamefont {Lumer},
  \citenamefont {Plotnik}, \citenamefont {Rechtsman},\ and\ \citenamefont
  {Segev}}]{Lumer2013}%
  \BibitemOpen
  \bibfield  {author} {\bibinfo {author} {\bibfnamefont {Y.}~\bibnamefont
  {Lumer}}, \bibinfo {author} {\bibfnamefont {Y.}~\bibnamefont {Plotnik}},
  \bibinfo {author} {\bibfnamefont {M.~C.}\ \bibnamefont {Rechtsman}},\ and\
  \bibinfo {author} {\bibfnamefont {M.}~\bibnamefont {Segev}},\ }\bibfield
  {title} {\bibinfo {title} {Self-localized states in photonic topological
  insulators},\ }\href {https://doi.org/10.1103/PhysRevLett.111.243905}
  {\bibfield  {journal} {\bibinfo  {journal} {Physical Review Letters}\
  }\textbf {\bibinfo {volume} {111}},\ \bibinfo {pages} {243905} (\bibinfo
  {year} {2013})}\BibitemShut {NoStop}%
\bibitem [{\citenamefont {Mukherjee}\ and\ \citenamefont
  {Rechtsman}(2020{\natexlab{a}})}]{Mukherjee2020a}%
  \BibitemOpen
  \bibfield  {author} {\bibinfo {author} {\bibfnamefont {S.}~\bibnamefont
  {Mukherjee}}\ and\ \bibinfo {author} {\bibfnamefont {M.~C.}\ \bibnamefont
  {Rechtsman}},\ }\bibfield  {title} {\bibinfo {title} {Observation of floquet
  solitons in a topological bandgap},\ }\href
  {https://doi.org/10.1126/science.aba8725} {\bibfield  {journal} {\bibinfo
  {journal} {Science}\ }\textbf {\bibinfo {volume} {368}},\ \bibinfo {pages}
  {856} (\bibinfo {year} {2020}{\natexlab{a}})}\BibitemShut {NoStop}%
\bibitem [{\citenamefont {Ablowitz}\ \emph {et~al.}(2014)\citenamefont
  {Ablowitz}, \citenamefont {Curtis},\ and\ \citenamefont {Ma}}]{Ablowitz2014}%
  \BibitemOpen
  \bibfield  {author} {\bibinfo {author} {\bibfnamefont {M.~J.}\ \bibnamefont
  {Ablowitz}}, \bibinfo {author} {\bibfnamefont {C.~W.}\ \bibnamefont
  {Curtis}},\ and\ \bibinfo {author} {\bibfnamefont {Y.-P.}\ \bibnamefont
  {Ma}},\ }\bibfield  {title} {\bibinfo {title} {Linear and nonlinear traveling
  edge waves in optical honeycomb lattices},\ }\href
  {https://doi.org/10.1103/PhysRevA.90.023813} {\bibfield  {journal} {\bibinfo
  {journal} {Physical Review A}\ }\textbf {\bibinfo {volume} {90}},\ \bibinfo
  {pages} {023813} (\bibinfo {year} {2014})}\BibitemShut {NoStop}%
\bibitem [{\citenamefont {Leykam}\ and\ \citenamefont
  {Chong}(2016)}]{Leykam2016}%
  \BibitemOpen
  \bibfield  {author} {\bibinfo {author} {\bibfnamefont {D.}~\bibnamefont
  {Leykam}}\ and\ \bibinfo {author} {\bibfnamefont {Y.~D.}\ \bibnamefont
  {Chong}},\ }\bibfield  {title} {\bibinfo {title} {Edge solitons in
  nonlinear-photonic topological insulators},\ }\href
  {https://doi.org/10.1103/PhysRevLett.117.143901} {\bibfield  {journal}
  {\bibinfo  {journal} {Physical Review Letters}\ }\textbf {\bibinfo {volume}
  {117}},\ \bibinfo {pages} {143901} (\bibinfo {year} {2016})}\BibitemShut
  {NoStop}%
\bibitem [{\citenamefont {Mukherjee}\ and\ \citenamefont
  {Rechtsman}(2020{\natexlab{b}})}]{Mukherjee2020b}%
  \BibitemOpen
  \bibfield  {author} {\bibinfo {author} {\bibfnamefont {S.}~\bibnamefont
  {Mukherjee}}\ and\ \bibinfo {author} {\bibfnamefont {M.~C.}\ \bibnamefont
  {Rechtsman}},\ }\bibfield  {title} {\bibinfo {title} {Observation of
  unidirectional soliton-like edge states in nonlinear floquet topological
  insulators},\ }\href@noop {} {\bibfield  {journal} {\bibinfo  {journal}
  {arXiv preprint arXiv:2010.11359}\ } (\bibinfo {year}
  {2020}{\natexlab{b}})}\BibitemShut {NoStop}%
\bibitem [{\citenamefont {Maczewsky}\ \emph {et~al.}(2020)\citenamefont
  {Maczewsky}, \citenamefont {Heinrich}, \citenamefont {Kremer}, \citenamefont
  {Ivanov}, \citenamefont {Ehrhardt}, \citenamefont {Martinez}, \citenamefont
  {Kartashov}, \citenamefont {Konotop}, \citenamefont {Torner}, \citenamefont
  {Bauer},\ and\ \citenamefont {Szameit}}]{Maczewsky2020}%
  \BibitemOpen
  \bibfield  {author} {\bibinfo {author} {\bibfnamefont {L.~J.}\ \bibnamefont
  {Maczewsky}}, \bibinfo {author} {\bibfnamefont {M.}~\bibnamefont {Heinrich}},
  \bibinfo {author} {\bibfnamefont {M.}~\bibnamefont {Kremer}}, \bibinfo
  {author} {\bibfnamefont {S.~K.}\ \bibnamefont {Ivanov}}, \bibinfo {author}
  {\bibfnamefont {M.}~\bibnamefont {Ehrhardt}}, \bibinfo {author}
  {\bibfnamefont {F.}~\bibnamefont {Martinez}}, \bibinfo {author}
  {\bibfnamefont {Y.~V.}\ \bibnamefont {Kartashov}}, \bibinfo {author}
  {\bibfnamefont {V.~V.}\ \bibnamefont {Konotop}}, \bibinfo {author}
  {\bibfnamefont {L.}~\bibnamefont {Torner}}, \bibinfo {author} {\bibfnamefont
  {D.}~\bibnamefont {Bauer}},\ and\ \bibinfo {author} {\bibfnamefont
  {A.}~\bibnamefont {Szameit}},\ }\bibfield  {title} {\bibinfo {title}
  {Nonlinearity-induced photonic topological insulator},\ }\href
  {https://doi.org/10.1126/science.abd2033} {\bibfield  {journal} {\bibinfo
  {journal} {Science}\ }\textbf {\bibinfo {volume} {370}},\ \bibinfo {pages}
  {701} (\bibinfo {year} {2020})}\BibitemShut {NoStop}%
\bibitem [{\citenamefont {Xia}\ \emph {et~al.}(2020)\citenamefont {Xia},
  \citenamefont {Jukić}, \citenamefont {Wang}, \citenamefont {Smirnova},
  \citenamefont {Smirnov}, \citenamefont {Tang}, \citenamefont {Song},
  \citenamefont {Szameit}, \citenamefont {Leykam}, \citenamefont {Xu},
  \citenamefont {Chen},\ and\ \citenamefont {Buljan}}]{Xia2020}%
  \BibitemOpen
  \bibfield  {author} {\bibinfo {author} {\bibfnamefont {S.}~\bibnamefont
  {Xia}}, \bibinfo {author} {\bibfnamefont {D.}~\bibnamefont {Jukić}},
  \bibinfo {author} {\bibfnamefont {N.}~\bibnamefont {Wang}}, \bibinfo {author}
  {\bibfnamefont {D.}~\bibnamefont {Smirnova}}, \bibinfo {author}
  {\bibfnamefont {L.}~\bibnamefont {Smirnov}}, \bibinfo {author} {\bibfnamefont
  {L.}~\bibnamefont {Tang}}, \bibinfo {author} {\bibfnamefont {D.}~\bibnamefont
  {Song}}, \bibinfo {author} {\bibfnamefont {A.}~\bibnamefont {Szameit}},
  \bibinfo {author} {\bibfnamefont {D.}~\bibnamefont {Leykam}}, \bibinfo
  {author} {\bibfnamefont {J.}~\bibnamefont {Xu}}, \bibinfo {author}
  {\bibfnamefont {Z.}~\bibnamefont {Chen}},\ and\ \bibinfo {author}
  {\bibfnamefont {H.}~\bibnamefont {Buljan}},\ }\bibfield  {title} {\bibinfo
  {title} {Nontrivial coupling of light into a defect: the interplay of
  nonlinearity and topology},\ }\href
  {https://doi.org/10.1038/s41377-020-00371-y} {\bibfield  {journal} {\bibinfo
  {journal} {Light: Science \& Applications}\ }\textbf {\bibinfo {volume}
  {9}},\ \bibinfo {pages} {147} (\bibinfo {year} {2020})}\BibitemShut {NoStop}%
\bibitem [{\citenamefont {Xia}\ \emph {et~al.}(2021)\citenamefont {Xia},
  \citenamefont {Kaltsas}, \citenamefont {Song}, \citenamefont {Komis},
  \citenamefont {Xu}, \citenamefont {Szameit}, \citenamefont {Buljan},
  \citenamefont {Makris},\ and\ \citenamefont {Chen}}]{Xia2021}%
  \BibitemOpen
  \bibfield  {author} {\bibinfo {author} {\bibfnamefont {S.}~\bibnamefont
  {Xia}}, \bibinfo {author} {\bibfnamefont {D.}~\bibnamefont {Kaltsas}},
  \bibinfo {author} {\bibfnamefont {D.}~\bibnamefont {Song}}, \bibinfo {author}
  {\bibfnamefont {I.}~\bibnamefont {Komis}}, \bibinfo {author} {\bibfnamefont
  {J.}~\bibnamefont {Xu}}, \bibinfo {author} {\bibfnamefont {A.}~\bibnamefont
  {Szameit}}, \bibinfo {author} {\bibfnamefont {H.}~\bibnamefont {Buljan}},
  \bibinfo {author} {\bibfnamefont {K.~G.}\ \bibnamefont {Makris}},\ and\
  \bibinfo {author} {\bibfnamefont {Z.}~\bibnamefont {Chen}},\ }\bibfield
  {title} {\bibinfo {title} {Nonlinear tuning of pt symmetry and non-hermitian
  topological states},\ }\href {https://doi.org/10.1126/science.abf6873}
  {\bibfield  {journal} {\bibinfo  {journal} {Science}\ }\textbf {\bibinfo
  {volume} {372}},\ \bibinfo {pages} {72} (\bibinfo {year} {2021})}\BibitemShut
  {NoStop}%
\bibitem [{\citenamefont {Thouless}(1983)}]{Thouless1983a}%
  \BibitemOpen
  \bibfield  {author} {\bibinfo {author} {\bibfnamefont {D.~J.}\ \bibnamefont
  {Thouless}},\ }\bibfield  {title} {\bibinfo {title} {Quantization of particle
  transport},\ }\href {https://doi.org/10.1103/PhysRevB.27.6083} {\bibfield
  {journal} {\bibinfo  {journal} {Physical Review B}\ }\textbf {\bibinfo
  {volume} {27}},\ \bibinfo {pages} {6083} (\bibinfo {year}
  {1983})}\BibitemShut {NoStop}%
\bibitem [{\citenamefont {Niu}\ and\ \citenamefont {Thouless}(1984)}]{Niu1984}%
  \BibitemOpen
  \bibfield  {author} {\bibinfo {author} {\bibfnamefont {Q.}~\bibnamefont
  {Niu}}\ and\ \bibinfo {author} {\bibfnamefont {D.}~\bibnamefont {Thouless}},\
  }\bibfield  {title} {\bibinfo {title} {Quantised adiabatic charge transport
  in the presence of substrate disorder and many-body interaction},\
  }\href@noop {} {\bibfield  {journal} {\bibinfo  {journal} {Journal of Physics
  A: Mathematical and General}\ }\textbf {\bibinfo {volume} {17}},\ \bibinfo
  {pages} {2453} (\bibinfo {year} {1984})}\BibitemShut {NoStop}%
\bibitem [{\citenamefont {Kraus}\ \emph {et~al.}(2012)\citenamefont {Kraus},
  \citenamefont {Lahini}, \citenamefont {Ringel}, \citenamefont {Verbin},\ and\
  \citenamefont {Zilberberg}}]{Kraus2012}%
  \BibitemOpen
  \bibfield  {author} {\bibinfo {author} {\bibfnamefont {Y.~E.}\ \bibnamefont
  {Kraus}}, \bibinfo {author} {\bibfnamefont {Y.}~\bibnamefont {Lahini}},
  \bibinfo {author} {\bibfnamefont {Z.}~\bibnamefont {Ringel}}, \bibinfo
  {author} {\bibfnamefont {M.}~\bibnamefont {Verbin}},\ and\ \bibinfo {author}
  {\bibfnamefont {O.}~\bibnamefont {Zilberberg}},\ }\bibfield  {title}
  {\bibinfo {title} {Topological states and adiabatic pumping in
  quasicrystals},\ }\href {https://doi.org/10.1103/PhysRevLett.109.106402}
  {\bibfield  {journal} {\bibinfo  {journal} {Physical Review Letters}\
  }\textbf {\bibinfo {volume} {109}},\ \bibinfo {pages} {106402} (\bibinfo
  {year} {2012})}\BibitemShut {NoStop}%
\bibitem [{\citenamefont {Lohse}\ \emph {et~al.}(2016)\citenamefont {Lohse},
  \citenamefont {Schweizer}, \citenamefont {Zilberberg}, \citenamefont
  {Aidelsburger},\ and\ \citenamefont {Bloch}}]{Lohse2016}%
  \BibitemOpen
  \bibfield  {author} {\bibinfo {author} {\bibfnamefont {M.}~\bibnamefont
  {Lohse}}, \bibinfo {author} {\bibfnamefont {C.}~\bibnamefont {Schweizer}},
  \bibinfo {author} {\bibfnamefont {O.}~\bibnamefont {Zilberberg}}, \bibinfo
  {author} {\bibfnamefont {M.}~\bibnamefont {Aidelsburger}},\ and\ \bibinfo
  {author} {\bibfnamefont {I.}~\bibnamefont {Bloch}},\ }\bibfield  {title}
  {\bibinfo {title} {A thouless quantum pump with ultracold bosonic atoms in an
  optical superlattice},\ }\href {https://doi.org/10.1038/nphys3584} {\bibfield
   {journal} {\bibinfo  {journal} {Nature Physics}\ }\textbf {\bibinfo {volume}
  {12}},\ \bibinfo {pages} {350} (\bibinfo {year} {2016})}\BibitemShut
  {NoStop}%
\bibitem [{\citenamefont {Nakajima}\ \emph {et~al.}(2016)\citenamefont
  {Nakajima}, \citenamefont {Tomita}, \citenamefont {Taie}, \citenamefont
  {Ichinose}, \citenamefont {Ozawa}, \citenamefont {Wang}, \citenamefont
  {Troyer},\ and\ \citenamefont {Takahashi}}]{Nakajima2016}%
  \BibitemOpen
  \bibfield  {author} {\bibinfo {author} {\bibfnamefont {S.}~\bibnamefont
  {Nakajima}}, \bibinfo {author} {\bibfnamefont {T.}~\bibnamefont {Tomita}},
  \bibinfo {author} {\bibfnamefont {S.}~\bibnamefont {Taie}}, \bibinfo {author}
  {\bibfnamefont {T.}~\bibnamefont {Ichinose}}, \bibinfo {author}
  {\bibfnamefont {H.}~\bibnamefont {Ozawa}}, \bibinfo {author} {\bibfnamefont
  {L.}~\bibnamefont {Wang}}, \bibinfo {author} {\bibfnamefont {M.}~\bibnamefont
  {Troyer}},\ and\ \bibinfo {author} {\bibfnamefont {Y.}~\bibnamefont
  {Takahashi}},\ }\bibfield  {title} {\bibinfo {title} {Topological thouless
  pumping of ultracold fermions},\ }\href {https://doi.org/10.1038/nphys3622}
  {\bibfield  {journal} {\bibinfo  {journal} {Nature Physics}\ }\textbf
  {\bibinfo {volume} {12}},\ \bibinfo {pages} {296} (\bibinfo {year}
  {2016})}\BibitemShut {NoStop}%
\bibitem [{\citenamefont {Lohse}\ \emph {et~al.}(2018)\citenamefont {Lohse},
  \citenamefont {Schweizer}, \citenamefont {Price}, \citenamefont
  {Zilberberg},\ and\ \citenamefont {Bloch}}]{Lohse2018}%
  \BibitemOpen
  \bibfield  {author} {\bibinfo {author} {\bibfnamefont {M.}~\bibnamefont
  {Lohse}}, \bibinfo {author} {\bibfnamefont {C.}~\bibnamefont {Schweizer}},
  \bibinfo {author} {\bibfnamefont {H.~M.}\ \bibnamefont {Price}}, \bibinfo
  {author} {\bibfnamefont {O.}~\bibnamefont {Zilberberg}},\ and\ \bibinfo
  {author} {\bibfnamefont {I.}~\bibnamefont {Bloch}},\ }\bibfield  {title}
  {\bibinfo {title} {Exploring 4d quantum hall physics with a 2d topological
  charge pump},\ }\href {https://doi.org/10.1038/nature25000} {\bibfield
  {journal} {\bibinfo  {journal} {Nature}\ }\textbf {\bibinfo {volume} {553}},\
  \bibinfo {pages} {55} (\bibinfo {year} {2018})}\BibitemShut {NoStop}%
\bibitem [{\citenamefont {Ma}\ \emph {et~al.}(2018)\citenamefont {Ma},
  \citenamefont {Zhou}, \citenamefont {Zhang}, \citenamefont {Li},
  \citenamefont {Cheng}, \citenamefont {Geng}, \citenamefont {Rong},
  \citenamefont {Shi}, \citenamefont {Gong},\ and\ \citenamefont
  {Du}}]{Wenchao2018}%
  \BibitemOpen
  \bibfield  {author} {\bibinfo {author} {\bibfnamefont {W.}~\bibnamefont
  {Ma}}, \bibinfo {author} {\bibfnamefont {L.}~\bibnamefont {Zhou}}, \bibinfo
  {author} {\bibfnamefont {Q.}~\bibnamefont {Zhang}}, \bibinfo {author}
  {\bibfnamefont {M.}~\bibnamefont {Li}}, \bibinfo {author} {\bibfnamefont
  {C.}~\bibnamefont {Cheng}}, \bibinfo {author} {\bibfnamefont
  {J.}~\bibnamefont {Geng}}, \bibinfo {author} {\bibfnamefont {X.}~\bibnamefont
  {Rong}}, \bibinfo {author} {\bibfnamefont {F.}~\bibnamefont {Shi}}, \bibinfo
  {author} {\bibfnamefont {J.}~\bibnamefont {Gong}},\ and\ \bibinfo {author}
  {\bibfnamefont {J.}~\bibnamefont {Du}},\ }\bibfield  {title} {\bibinfo
  {title} {Experimental observation of a generalized thouless pump with a
  single spin},\ }\href {https://doi.org/10.1103/PhysRevLett.120.120501}
  {\bibfield  {journal} {\bibinfo  {journal} {Physical Review Letters}\
  }\textbf {\bibinfo {volume} {120}},\ \bibinfo {pages} {120501} (\bibinfo
  {year} {2018})}\BibitemShut {NoStop}%
\bibitem [{\citenamefont {Fedorova}\ \emph {et~al.}(2020)\citenamefont
  {Fedorova}, \citenamefont {Qiu}, \citenamefont {Linden},\ and\ \citenamefont
  {Kroha}}]{Fedorova2020}%
  \BibitemOpen
  \bibfield  {author} {\bibinfo {author} {\bibfnamefont {Z.}~\bibnamefont
  {Fedorova}}, \bibinfo {author} {\bibfnamefont {H.}~\bibnamefont {Qiu}},
  \bibinfo {author} {\bibfnamefont {S.}~\bibnamefont {Linden}},\ and\ \bibinfo
  {author} {\bibfnamefont {J.}~\bibnamefont {Kroha}},\ }\bibfield  {title}
  {\bibinfo {title} {Observation of topological transport quantization by
  dissipation in fast thouless pumps},\ }\href
  {https://doi.org/10.1038/s41467-020-17510-z} {\bibfield  {journal} {\bibinfo
  {journal} {Nature Communications}\ }\textbf {\bibinfo {volume} {11}},\
  \bibinfo {pages} {3758} (\bibinfo {year} {2020})}\BibitemShut {NoStop}%
\bibitem [{\citenamefont {Cerjan}\ \emph {et~al.}(2020)\citenamefont {Cerjan},
  \citenamefont {Wang}, \citenamefont {Huang}, \citenamefont {Chen},\ and\
  \citenamefont {Rechtsman}}]{Cerjan2020Thouless}%
  \BibitemOpen
  \bibfield  {author} {\bibinfo {author} {\bibfnamefont {A.}~\bibnamefont
  {Cerjan}}, \bibinfo {author} {\bibfnamefont {M.}~\bibnamefont {Wang}},
  \bibinfo {author} {\bibfnamefont {S.}~\bibnamefont {Huang}}, \bibinfo
  {author} {\bibfnamefont {K.~P.}\ \bibnamefont {Chen}},\ and\ \bibinfo
  {author} {\bibfnamefont {M.~C.}\ \bibnamefont {Rechtsman}},\ }\bibfield
  {title} {\bibinfo {title} {Thouless pumping in disordered photonic systems},\
  }\href {https://doi.org/10.1038/s41377-020-00408-2} {\bibfield  {journal}
  {\bibinfo  {journal} {Light: Science \& Applications}\ }\textbf {\bibinfo
  {volume} {9}},\ \bibinfo {pages} {178} (\bibinfo {year} {2020})}\BibitemShut
  {NoStop}%
\bibitem [{\citenamefont {Grinberg}\ \emph {et~al.}(2020)\citenamefont
  {Grinberg}, \citenamefont {Lin}, \citenamefont {Harris}, \citenamefont
  {Benalcazar}, \citenamefont {Peterson}, \citenamefont {Hughes},\ and\
  \citenamefont {Bahl}}]{Grinberg2020}%
  \BibitemOpen
  \bibfield  {author} {\bibinfo {author} {\bibfnamefont {I.~H.}\ \bibnamefont
  {Grinberg}}, \bibinfo {author} {\bibfnamefont {M.}~\bibnamefont {Lin}},
  \bibinfo {author} {\bibfnamefont {C.}~\bibnamefont {Harris}}, \bibinfo
  {author} {\bibfnamefont {W.~A.}\ \bibnamefont {Benalcazar}}, \bibinfo
  {author} {\bibfnamefont {C.~W.}\ \bibnamefont {Peterson}}, \bibinfo {author}
  {\bibfnamefont {T.~L.}\ \bibnamefont {Hughes}},\ and\ \bibinfo {author}
  {\bibfnamefont {G.}~\bibnamefont {Bahl}},\ }\bibfield  {title} {\bibinfo
  {title} {Robust temporal pumping in a magneto-mechanical topological
  insulator},\ }\href {https://doi.org/10.1038/s41467-020-14804-0} {\bibfield
  {journal} {\bibinfo  {journal} {Nature Communications}\ }\textbf {\bibinfo
  {volume} {11}},\ \bibinfo {pages} {974} (\bibinfo {year} {2020})}\BibitemShut
  {NoStop}%
\bibitem [{\citenamefont {Jürgensen}\ \emph {et~al.}(2021)\citenamefont
  {Jürgensen}, \citenamefont {Mukherjee},\ and\ \citenamefont
  {Rechtsman}}]{Juergensen2021}%
  \BibitemOpen
  \bibfield  {author} {\bibinfo {author} {\bibfnamefont {M.}~\bibnamefont
  {Jürgensen}}, \bibinfo {author} {\bibfnamefont {S.}~\bibnamefont
  {Mukherjee}},\ and\ \bibinfo {author} {\bibfnamefont {M.~C.}\ \bibnamefont
  {Rechtsman}},\ }\bibfield  {title} {\bibinfo {title} {Quantized nonlinear
  thouless pumping},\ }\href {https://doi.org/10.1038/s41586-021-03688-9}
  {\bibfield  {journal} {\bibinfo  {journal} {Nature}\ }\textbf {\bibinfo
  {volume} {596}},\ \bibinfo {pages} {63} (\bibinfo {year} {2021})}\BibitemShut
  {NoStop}%
\bibitem [{\citenamefont {Christodoulides}\ and\ \citenamefont
  {Joseph}(1988)}]{Christodoulides1988}%
  \BibitemOpen
  \bibfield  {author} {\bibinfo {author} {\bibfnamefont {D.~N.}\ \bibnamefont
  {Christodoulides}}\ and\ \bibinfo {author} {\bibfnamefont {R.~I.}\
  \bibnamefont {Joseph}},\ }\bibfield  {title} {\bibinfo {title} {Discrete
  self-focusing in nonlinear arrays of coupled waveguides},\ }\href
  {https://doi.org/10.1364/OL.13.000794} {\bibfield  {journal} {\bibinfo
  {journal} {Optics Letters}\ }\textbf {\bibinfo {volume} {13}},\ \bibinfo
  {pages} {794} (\bibinfo {year} {1988})}\BibitemShut {NoStop}%
\bibitem [{\citenamefont {Eisenberg}\ \emph {et~al.}(1998)\citenamefont
  {Eisenberg}, \citenamefont {Silberberg}, \citenamefont {Morandotti},
  \citenamefont {Boyd},\ and\ \citenamefont
  {Aitchison}}]{eisenberg1998discrete}%
  \BibitemOpen
  \bibfield  {author} {\bibinfo {author} {\bibfnamefont {H.}~\bibnamefont
  {Eisenberg}}, \bibinfo {author} {\bibfnamefont {Y.}~\bibnamefont
  {Silberberg}}, \bibinfo {author} {\bibfnamefont {R.}~\bibnamefont
  {Morandotti}}, \bibinfo {author} {\bibfnamefont {A.}~\bibnamefont {Boyd}},\
  and\ \bibinfo {author} {\bibfnamefont {J.}~\bibnamefont {Aitchison}},\
  }\bibfield  {title} {\bibinfo {title} {Discrete spatial optical solitons in
  waveguide arrays},\ }\href@noop {} {\bibfield  {journal} {\bibinfo  {journal}
  {Physical Review Letters}\ }\textbf {\bibinfo {volume} {81}},\ \bibinfo
  {pages} {3383} (\bibinfo {year} {1998})}\BibitemShut {NoStop}%
\bibitem [{\citenamefont {Fleischer}\ \emph {et~al.}(2003)\citenamefont
  {Fleischer}, \citenamefont {Segev}, \citenamefont {Efremidis},\ and\
  \citenamefont {Christodoulides}}]{fleischer2003observation}%
  \BibitemOpen
  \bibfield  {author} {\bibinfo {author} {\bibfnamefont {J.~W.}\ \bibnamefont
  {Fleischer}}, \bibinfo {author} {\bibfnamefont {M.}~\bibnamefont {Segev}},
  \bibinfo {author} {\bibfnamefont {N.~K.}\ \bibnamefont {Efremidis}},\ and\
  \bibinfo {author} {\bibfnamefont {D.~N.}\ \bibnamefont {Christodoulides}},\
  }\bibfield  {title} {\bibinfo {title} {Observation of two-dimensional
  discrete solitons in optically induced nonlinear photonic lattices},\
  }\href@noop {} {\bibfield  {journal} {\bibinfo  {journal} {Nature}\ }\textbf
  {\bibinfo {volume} {422}},\ \bibinfo {pages} {147} (\bibinfo {year}
  {2003})}\BibitemShut {NoStop}%
\bibitem [{\citenamefont {Christodoulides}\ \emph {et~al.}(2003)\citenamefont
  {Christodoulides}, \citenamefont {Lederer},\ and\ \citenamefont
  {Silberberg}}]{christodoulides2003discretizing}%
  \BibitemOpen
  \bibfield  {author} {\bibinfo {author} {\bibfnamefont {D.~N.}\ \bibnamefont
  {Christodoulides}}, \bibinfo {author} {\bibfnamefont {F.}~\bibnamefont
  {Lederer}},\ and\ \bibinfo {author} {\bibfnamefont {Y.}~\bibnamefont
  {Silberberg}},\ }\bibfield  {title} {\bibinfo {title} {Discretizing light
  behaviour in linear and nonlinear waveguide lattices},\ }\href
  {https://doi.org/10.1038/nature01936} {\bibfield  {journal} {\bibinfo
  {journal} {Nature}\ }\textbf {\bibinfo {volume} {424}},\ \bibinfo {pages}
  {817} (\bibinfo {year} {2003})}\BibitemShut {NoStop}%
\bibitem [{\citenamefont {Kivshar}\ and\ \citenamefont
  {Agrawal}(2003)}]{Kivshar2003}%
  \BibitemOpen
  \bibfield  {author} {\bibinfo {author} {\bibfnamefont {Y.~S.}\ \bibnamefont
  {Kivshar}}\ and\ \bibinfo {author} {\bibfnamefont {G.~P.}\ \bibnamefont
  {Agrawal}},\ }\href@noop {} {\emph {\bibinfo {title} {Optical solitons: from
  fibers to photonic crystals}}}\ (\bibinfo  {publisher} {Academic press},\
  \bibinfo {year} {2003})\BibitemShut {NoStop}%
\bibitem [{\citenamefont {Lederer}\ \emph {et~al.}(2008)\citenamefont
  {Lederer}, \citenamefont {Stegeman}, \citenamefont {Christodoulides},
  \citenamefont {Assanto}, \citenamefont {Segev},\ and\ \citenamefont
  {Silberberg}}]{lederer2008discrete}%
  \BibitemOpen
  \bibfield  {author} {\bibinfo {author} {\bibfnamefont {F.}~\bibnamefont
  {Lederer}}, \bibinfo {author} {\bibfnamefont {G.~I.}\ \bibnamefont
  {Stegeman}}, \bibinfo {author} {\bibfnamefont {D.~N.}\ \bibnamefont
  {Christodoulides}}, \bibinfo {author} {\bibfnamefont {G.}~\bibnamefont
  {Assanto}}, \bibinfo {author} {\bibfnamefont {M.}~\bibnamefont {Segev}},\
  and\ \bibinfo {author} {\bibfnamefont {Y.}~\bibnamefont {Silberberg}},\
  }\bibfield  {title} {\bibinfo {title} {Discrete solitons in optics},\ }\href
  {https://doi.org/https://doi.org/10.1016/j.physrep.2008.04.004} {\bibfield
  {journal} {\bibinfo  {journal} {Physics Reports}\ }\textbf {\bibinfo {volume}
  {463}},\ \bibinfo {pages} {1} (\bibinfo {year} {2008})}\BibitemShut {NoStop}%
\bibitem [{\citenamefont {Kevrekidis}(2009)}]{Kevrekidis2009}%
  \BibitemOpen
  \bibfield  {author} {\bibinfo {author} {\bibfnamefont {P.~G.}\ \bibnamefont
  {Kevrekidis}},\ }\href@noop {} {\emph {\bibinfo {title} {The discrete
  nonlinear Schrödinger equation: mathematical analysis, numerical
  computations and physical perspectives}}},\ Vol.\ \bibinfo {volume} {232}\
  (\bibinfo  {publisher} {Springer Science \& Business Media},\ \bibinfo {year}
  {2009})\BibitemShut {NoStop}%
\bibitem [{\citenamefont {Dalfovo}\ \emph {et~al.}(1999)\citenamefont
  {Dalfovo}, \citenamefont {Giorgini}, \citenamefont {Pitaevskii},\ and\
  \citenamefont {Stringari}}]{Dalfavo1999}%
  \BibitemOpen
  \bibfield  {author} {\bibinfo {author} {\bibfnamefont {F.}~\bibnamefont
  {Dalfovo}}, \bibinfo {author} {\bibfnamefont {S.}~\bibnamefont {Giorgini}},
  \bibinfo {author} {\bibfnamefont {L.~P.}\ \bibnamefont {Pitaevskii}},\ and\
  \bibinfo {author} {\bibfnamefont {S.}~\bibnamefont {Stringari}},\ }\bibfield
  {title} {\bibinfo {title} {Theory of bose-einstein condensation in trapped
  gases},\ }\href {https://doi.org/10.1103/RevModPhys.71.463} {\bibfield
  {journal} {\bibinfo  {journal} {Reviews of Modern Physics}\ }\textbf
  {\bibinfo {volume} {71}},\ \bibinfo {pages} {463} (\bibinfo {year}
  {1999})}\BibitemShut {NoStop}%
\bibitem [{\citenamefont {Pitaevskii}\ and\ \citenamefont
  {Stringari}(2016)}]{Pitaevskii2016Bose}%
  \BibitemOpen
  \bibfield  {author} {\bibinfo {author} {\bibfnamefont {L.}~\bibnamefont
  {Pitaevskii}}\ and\ \bibinfo {author} {\bibfnamefont {S.}~\bibnamefont
  {Stringari}},\ }\href@noop {} {\emph {\bibinfo {title} {Bose-Einstein
  condensation and superfluidity}}},\ Vol.\ \bibinfo {volume} {164}\ (\bibinfo
  {publisher} {Oxford University Press},\ \bibinfo {year} {2016})\BibitemShut
  {NoStop}%
\bibitem [{\citenamefont {Donley}\ \emph {et~al.}(2001)\citenamefont {Donley},
  \citenamefont {Claussen}, \citenamefont {Cornish}, \citenamefont {Roberts},
  \citenamefont {Cornell},\ and\ \citenamefont {Wieman}}]{Donley2001}%
  \BibitemOpen
  \bibfield  {author} {\bibinfo {author} {\bibfnamefont {E.~A.}\ \bibnamefont
  {Donley}}, \bibinfo {author} {\bibfnamefont {N.~R.}\ \bibnamefont
  {Claussen}}, \bibinfo {author} {\bibfnamefont {S.~L.}\ \bibnamefont
  {Cornish}}, \bibinfo {author} {\bibfnamefont {J.~L.}\ \bibnamefont
  {Roberts}}, \bibinfo {author} {\bibfnamefont {E.~A.}\ \bibnamefont
  {Cornell}},\ and\ \bibinfo {author} {\bibfnamefont {C.~E.}\ \bibnamefont
  {Wieman}},\ }\bibfield  {title} {\bibinfo {title} {Dynamics of collapsing and
  exploding bose–einstein condensates},\ }\href
  {https://doi.org/10.1038/35085500} {\bibfield  {journal} {\bibinfo  {journal}
  {Nature}\ }\textbf {\bibinfo {volume} {412}},\ \bibinfo {pages} {295}
  (\bibinfo {year} {2001})}\BibitemShut {NoStop}%
\bibitem [{\citenamefont {Abo-Shaeer}\ \emph {et~al.}(2001)\citenamefont
  {Abo-Shaeer}, \citenamefont {Raman}, \citenamefont {Vogels},\ and\
  \citenamefont {Ketterle}}]{Abo-Shaeer2001Observation}%
  \BibitemOpen
  \bibfield  {author} {\bibinfo {author} {\bibfnamefont {J.~R.}\ \bibnamefont
  {Abo-Shaeer}}, \bibinfo {author} {\bibfnamefont {C.}~\bibnamefont {Raman}},
  \bibinfo {author} {\bibfnamefont {J.~M.}\ \bibnamefont {Vogels}},\ and\
  \bibinfo {author} {\bibfnamefont {W.}~\bibnamefont {Ketterle}},\ }\bibfield
  {title} {\bibinfo {title} {Observation of vortex lattices in bose-einstein
  condensates},\ }\href {https://doi.org/doi:10.1126/science.1060182}
  {\bibfield  {journal} {\bibinfo  {journal} {Science}\ }\textbf {\bibinfo
  {volume} {292}},\ \bibinfo {pages} {476} (\bibinfo {year}
  {2001})}\BibitemShut {NoStop}%
\bibitem [{\citenamefont {Deng}\ \emph {et~al.}(2002)\citenamefont {Deng},
  \citenamefont {Weihs}, \citenamefont {Santori}, \citenamefont {Bloch},\ and\
  \citenamefont {Yamamoto}}]{Deng2002}%
  \BibitemOpen
  \bibfield  {author} {\bibinfo {author} {\bibfnamefont {H.}~\bibnamefont
  {Deng}}, \bibinfo {author} {\bibfnamefont {G.}~\bibnamefont {Weihs}},
  \bibinfo {author} {\bibfnamefont {C.}~\bibnamefont {Santori}}, \bibinfo
  {author} {\bibfnamefont {J.}~\bibnamefont {Bloch}},\ and\ \bibinfo {author}
  {\bibfnamefont {Y.}~\bibnamefont {Yamamoto}},\ }\bibfield  {title} {\bibinfo
  {title} {Condensation of semiconductor microcavity exciton polaritons},\
  }\href {https://doi.org/doi:10.1126/science.1074464} {\bibfield  {journal}
  {\bibinfo  {journal} {Science}\ }\textbf {\bibinfo {volume} {298}},\ \bibinfo
  {pages} {199} (\bibinfo {year} {2002})}\BibitemShut {NoStop}%
\bibitem [{\citenamefont {Kasprzak}\ \emph {et~al.}(2006)\citenamefont
  {Kasprzak}, \citenamefont {Richard}, \citenamefont {Kundermann},
  \citenamefont {Baas}, \citenamefont {Jeambrun}, \citenamefont {Keeling},
  \citenamefont {Marchetti}, \citenamefont {Szymańska}, \citenamefont
  {André}, \citenamefont {Staehli}, \citenamefont {Savona}, \citenamefont
  {Littlewood}, \citenamefont {Deveaud},\ and\ \citenamefont
  {Dang}}]{Kasprzak2006}%
  \BibitemOpen
  \bibfield  {author} {\bibinfo {author} {\bibfnamefont {J.}~\bibnamefont
  {Kasprzak}}, \bibinfo {author} {\bibfnamefont {M.}~\bibnamefont {Richard}},
  \bibinfo {author} {\bibfnamefont {S.}~\bibnamefont {Kundermann}}, \bibinfo
  {author} {\bibfnamefont {A.}~\bibnamefont {Baas}}, \bibinfo {author}
  {\bibfnamefont {P.}~\bibnamefont {Jeambrun}}, \bibinfo {author}
  {\bibfnamefont {J.~M.~J.}\ \bibnamefont {Keeling}}, \bibinfo {author}
  {\bibfnamefont {F.~M.}\ \bibnamefont {Marchetti}}, \bibinfo {author}
  {\bibfnamefont {M.~H.}\ \bibnamefont {Szymańska}}, \bibinfo {author}
  {\bibfnamefont {R.}~\bibnamefont {André}}, \bibinfo {author} {\bibfnamefont
  {J.~L.}\ \bibnamefont {Staehli}}, \bibinfo {author} {\bibfnamefont
  {V.}~\bibnamefont {Savona}}, \bibinfo {author} {\bibfnamefont {P.~B.}\
  \bibnamefont {Littlewood}}, \bibinfo {author} {\bibfnamefont
  {B.}~\bibnamefont {Deveaud}},\ and\ \bibinfo {author} {\bibfnamefont {L.~S.}\
  \bibnamefont {Dang}},\ }\bibfield  {title} {\bibinfo {title} {Bose–einstein
  condensation of exciton polaritons},\ }\href
  {https://doi.org/10.1038/nature05131} {\bibfield  {journal} {\bibinfo
  {journal} {Nature}\ }\textbf {\bibinfo {volume} {443}},\ \bibinfo {pages}
  {409} (\bibinfo {year} {2006})}\BibitemShut {NoStop}%
\bibitem [{\citenamefont {Balili}\ \emph {et~al.}(2007)\citenamefont {Balili},
  \citenamefont {Hartwell}, \citenamefont {Snoke}, \citenamefont {Pfeiffer},\
  and\ \citenamefont {West}}]{Balili2007}%
  \BibitemOpen
  \bibfield  {author} {\bibinfo {author} {\bibfnamefont {R.}~\bibnamefont
  {Balili}}, \bibinfo {author} {\bibfnamefont {V.}~\bibnamefont {Hartwell}},
  \bibinfo {author} {\bibfnamefont {D.}~\bibnamefont {Snoke}}, \bibinfo
  {author} {\bibfnamefont {L.}~\bibnamefont {Pfeiffer}},\ and\ \bibinfo
  {author} {\bibfnamefont {K.}~\bibnamefont {West}},\ }\bibfield  {title}
  {\bibinfo {title} {Bose-einstein condensation of microcavity polaritons in a
  trap},\ }\href {https://doi.org/doi:10.1126/science.1140990} {\bibfield
  {journal} {\bibinfo  {journal} {Science}\ }\textbf {\bibinfo {volume}
  {316}},\ \bibinfo {pages} {1007} (\bibinfo {year} {2007})}\BibitemShut
  {NoStop}%
\bibitem [{\citenamefont {Harper}(1955)}]{Harper1955}%
  \BibitemOpen
  \bibfield  {author} {\bibinfo {author} {\bibfnamefont {P.~G.}\ \bibnamefont
  {Harper}},\ }\bibfield  {title} {\bibinfo {title} {Single band motion of
  conduction electrons in a uniform magnetic field},\ }\href@noop {} {\bibfield
   {journal} {\bibinfo  {journal} {Proceedings of the Physical Society. Section
  A}\ }\textbf {\bibinfo {volume} {68}},\ \bibinfo {pages} {874} (\bibinfo
  {year} {1955})}\BibitemShut {NoStop}%
\bibitem [{\citenamefont {Aubry}\ and\ \citenamefont
  {André}(1980)}]{Aubry1980}%
  \BibitemOpen
  \bibfield  {author} {\bibinfo {author} {\bibfnamefont {S.}~\bibnamefont
  {Aubry}}\ and\ \bibinfo {author} {\bibfnamefont {G.}~\bibnamefont {André}},\
  }\bibfield  {title} {\bibinfo {title} {Analyticity breaking and anderson
  localization in incommensurate lattices},\ }\href@noop {} {\bibfield
  {journal} {\bibinfo  {journal} {Ann. Israel Phys. Soc}\ }\textbf {\bibinfo
  {volume} {3}},\ \bibinfo {pages} {18} (\bibinfo {year} {1980})}\BibitemShut
  {NoStop}%
\bibitem [{\citenamefont {Ke}\ \emph {et~al.}(2016)\citenamefont {Ke},
  \citenamefont {Qin}, \citenamefont {Mei}, \citenamefont {Zhong},
  \citenamefont {Kivshar},\ and\ \citenamefont {Lee}}]{Ke2016}%
  \BibitemOpen
  \bibfield  {author} {\bibinfo {author} {\bibfnamefont {Y.}~\bibnamefont
  {Ke}}, \bibinfo {author} {\bibfnamefont {X.}~\bibnamefont {Qin}}, \bibinfo
  {author} {\bibfnamefont {F.}~\bibnamefont {Mei}}, \bibinfo {author}
  {\bibfnamefont {H.}~\bibnamefont {Zhong}}, \bibinfo {author} {\bibfnamefont
  {Y.~S.}\ \bibnamefont {Kivshar}},\ and\ \bibinfo {author} {\bibfnamefont
  {C.}~\bibnamefont {Lee}},\ }\bibfield  {title} {\bibinfo {title} {Topological
  phase transitions and thouless pumping of light in photonic waveguide
  arrays},\ }\href {https://doi.org/https://doi.org/10.1002/lpor.201600119}
  {\bibfield  {journal} {\bibinfo  {journal} {Laser \& Photonics Reviews}\
  }\textbf {\bibinfo {volume} {10}},\ \bibinfo {pages} {995} (\bibinfo {year}
  {2016})}\BibitemShut {NoStop}%
\bibitem [{\citenamefont {Askaryan}(1962)}]{Askaryan1962}%
  \BibitemOpen
  \bibfield  {author} {\bibinfo {author} {\bibfnamefont {G.}~\bibnamefont
  {Askaryan}},\ }\bibfield  {title} {\bibinfo {title} {Effect of the gradient
  of a strong electromagnetic ray on electrons and atoms},\ }\href@noop {}
  {\bibfield  {journal} {\bibinfo  {journal} {Zhur. Eksptl'. i Teoret. Fiz.}\
  }\textbf {\bibinfo {volume} {42}} (\bibinfo {year} {1962})}\BibitemShut
  {NoStop}%
\bibitem [{\citenamefont {Chiao}\ \emph {et~al.}(1964)\citenamefont {Chiao},
  \citenamefont {Garmire},\ and\ \citenamefont {Townes}}]{Chiao1964}%
  \BibitemOpen
  \bibfield  {author} {\bibinfo {author} {\bibfnamefont {R.~Y.}\ \bibnamefont
  {Chiao}}, \bibinfo {author} {\bibfnamefont {E.}~\bibnamefont {Garmire}},\
  and\ \bibinfo {author} {\bibfnamefont {C.~H.}\ \bibnamefont {Townes}},\
  }\bibfield  {title} {\bibinfo {title} {Self-trapping of optical beams},\
  }\href {https://doi.org/10.1103/PhysRevLett.13.479} {\bibfield  {journal}
  {\bibinfo  {journal} {Physical Review Letters}\ }\textbf {\bibinfo {volume}
  {13}},\ \bibinfo {pages} {479} (\bibinfo {year} {1964})}\BibitemShut
  {NoStop}%
\bibitem [{\citenamefont {Ablowitz}\ and\ \citenamefont
  {Segur}(1981)}]{Ablowitz1981}%
  \BibitemOpen
  \bibfield  {author} {\bibinfo {author} {\bibfnamefont {M.~J.}\ \bibnamefont
  {Ablowitz}}\ and\ \bibinfo {author} {\bibfnamefont {H.}~\bibnamefont
  {Segur}},\ }\href@noop {} {\emph {\bibinfo {title} {Solitons and the inverse
  scattering transform}}}\ (\bibinfo  {publisher} {SIAM},\ \bibinfo {year}
  {1981})\BibitemShut {NoStop}%
\bibitem [{\citenamefont {Stegeman}\ and\ \citenamefont
  {Segev}(1999)}]{Stegeman1999}%
  \BibitemOpen
  \bibfield  {author} {\bibinfo {author} {\bibfnamefont {G.~I.}\ \bibnamefont
  {Stegeman}}\ and\ \bibinfo {author} {\bibfnamefont {M.}~\bibnamefont
  {Segev}},\ }\bibfield  {title} {\bibinfo {title} {Optical spatial solitons
  and their interactions: Universality and diversity},\ }\href
  {https://doi.org/10.1126/science.286.5444.1518} {\bibfield  {journal}
  {\bibinfo  {journal} {Science}\ }\textbf {\bibinfo {volume} {286}},\ \bibinfo
  {pages} {1518} (\bibinfo {year} {1999})}\BibitemShut {NoStop}%
\bibitem [{\citenamefont {Jürgensen}\ and\ \citenamefont
  {Rechtsman}(2021)}]{Juergensen2021TheChern}%
  \BibitemOpen
  \bibfield  {author} {\bibinfo {author} {\bibfnamefont {M.}~\bibnamefont
  {Jürgensen}}\ and\ \bibinfo {author} {\bibfnamefont {M.~C.}\ \bibnamefont
  {Rechtsman}},\ }\bibfield  {title} {\bibinfo {title} {The chern number
  governs soliton motion in nonlinear thouless pumps},\ }\href@noop {}
  {\bibfield  {journal} {\bibinfo  {journal} {arXiv preprint arXiv:2110.08696}\
  } (\bibinfo {year} {2021})}\BibitemShut {NoStop}%
\bibitem [{\citenamefont {Mostaan}\ \emph {et~al.}(2021)\citenamefont
  {Mostaan}, \citenamefont {Grusdt},\ and\ \citenamefont
  {Goldman}}]{Mostaan2021}%
  \BibitemOpen
  \bibfield  {author} {\bibinfo {author} {\bibfnamefont {N.}~\bibnamefont
  {Mostaan}}, \bibinfo {author} {\bibfnamefont {F.}~\bibnamefont {Grusdt}},\
  and\ \bibinfo {author} {\bibfnamefont {N.}~\bibnamefont {Goldman}},\
  }\bibfield  {title} {\bibinfo {title} {Quantized transport of solitons in
  nonlinear thouless pumps: From wannier drags to topological polarons},\
  }\href@noop {} {\bibfield  {journal} {\bibinfo  {journal} {arXiv preprint
  arXiv:2110.13075}\ } (\bibinfo {year} {2021})}\BibitemShut {NoStop}%
\bibitem [{\citenamefont {Davis}\ \emph {et~al.}(1996)\citenamefont {Davis},
  \citenamefont {Miura}, \citenamefont {Sugimoto},\ and\ \citenamefont
  {Hirao}}]{Miura1996}%
  \BibitemOpen
  \bibfield  {author} {\bibinfo {author} {\bibfnamefont {K.~M.}\ \bibnamefont
  {Davis}}, \bibinfo {author} {\bibfnamefont {K.}~\bibnamefont {Miura}},
  \bibinfo {author} {\bibfnamefont {N.}~\bibnamefont {Sugimoto}},\ and\
  \bibinfo {author} {\bibfnamefont {K.}~\bibnamefont {Hirao}},\ }\bibfield
  {title} {\bibinfo {title} {Writing waveguides in glass with a femtosecond
  laser},\ }\href@noop {} {\bibfield  {journal} {\bibinfo  {journal} {Optics
  letters}\ }\textbf {\bibinfo {volume} {21}},\ \bibinfo {pages} {1729}
  (\bibinfo {year} {1996})}\BibitemShut {NoStop}%
\bibitem [{\citenamefont {Szameit}\ and\ \citenamefont
  {Nolte}(2010)}]{Szameit2010}%
  \BibitemOpen
  \bibfield  {author} {\bibinfo {author} {\bibfnamefont {A.}~\bibnamefont
  {Szameit}}\ and\ \bibinfo {author} {\bibfnamefont {S.}~\bibnamefont
  {Nolte}},\ }\bibfield  {title} {\bibinfo {title} {Discrete optics in
  femtosecond-laser-written photonic structures},\ }\href
  {https://doi.org/10.1088/0953-4075/43/16/163001} {\bibfield  {journal}
  {\bibinfo  {journal} {Journal of Physics B: Atomic, Molecular and Optical
  Physics}\ }\textbf {\bibinfo {volume} {43}},\ \bibinfo {pages} {163001}
  (\bibinfo {year} {2010})}\BibitemShut {NoStop}%
\bibitem [{\citenamefont {Fu}\ \emph {et~al.}(2021)\citenamefont {Fu},
  \citenamefont {Wang}, \citenamefont {Kartashov}, \citenamefont {Konotop},\
  and\ \citenamefont {Ye}}]{Fu2021}%
  \BibitemOpen
  \bibfield  {author} {\bibinfo {author} {\bibfnamefont {Q.}~\bibnamefont
  {Fu}}, \bibinfo {author} {\bibfnamefont {P.}~\bibnamefont {Wang}}, \bibinfo
  {author} {\bibfnamefont {Y.~V.}\ \bibnamefont {Kartashov}}, \bibinfo {author}
  {\bibfnamefont {V.~V.}\ \bibnamefont {Konotop}},\ and\ \bibinfo {author}
  {\bibfnamefont {F.}~\bibnamefont {Ye}},\ }\bibfield  {title} {\bibinfo
  {title} {Nonlinear thouless pumping: solitons and transport breakdown},\
  }\href@noop {} {\bibfield  {journal} {\bibinfo  {journal} {arXiv preprint
  arXiv:2110.14472}\ } (\bibinfo {year} {2021})}\BibitemShut {NoStop}%
\end{thebibliography}%


\begin{thebibliography}{4}%
\makeatletter
\providecommand \@ifxundefined [1]{%
 \@ifx{#1\undefined}
}%
\providecommand \@ifnum [1]{%
 \ifnum #1\expandafter \@firstoftwo
 \else \expandafter \@secondoftwo
 \fi
}%
\providecommand \@ifx [1]{%
 \ifx #1\expandafter \@firstoftwo
 \else \expandafter \@secondoftwo
 \fi
}%
\providecommand \natexlab [1]{#1}%
\providecommand \enquote  [1]{``#1''}%
\providecommand \bibnamefont  [1]{#1}%
\providecommand \bibfnamefont [1]{#1}%
\providecommand \citenamefont [1]{#1}%
\providecommand \href@noop [0]{\@secondoftwo}%
\providecommand \href [0]{\begingroup \@sanitize@url \@href}%
\providecommand \@href[1]{\@@startlink{#1}\@@href}%
\providecommand \@@href[1]{\endgroup#1\@@endlink}%
\providecommand \@sanitize@url [0]{\catcode `\\12\catcode `\$12\catcode
  `\&12\catcode `\#12\catcode `\^12\catcode `\_12\catcode `\%12\relax}%
\providecommand \@@startlink[1]{}%
\providecommand \@@endlink[0]{}%
\providecommand \url  [0]{\begingroup\@sanitize@url \@url }%
\providecommand \@url [1]{\endgroup\@href {#1}{\urlprefix }}%
\providecommand \urlprefix  [0]{URL }%
\providecommand \Eprint [0]{\href }%
\providecommand \doibase [0]{https://doi.org/}%
\providecommand \selectlanguage [0]{\@gobble}%
\providecommand \bibinfo  [0]{\@secondoftwo}%
\providecommand \bibfield  [0]{\@secondoftwo}%
\providecommand \translation [1]{[#1]}%
\providecommand \BibitemOpen [0]{}%
\providecommand \bibitemStop [0]{}%
\providecommand \bibitemNoStop [0]{.\EOS\space}%
\providecommand \EOS [0]{\spacefactor3000\relax}%
\providecommand \BibitemShut  [1]{\csname bibitem#1\endcsname}%
\let\auto@bib@innerbib\@empty
\bibitem [{\citenamefont {Davis}\ \emph {et~al.}(1996)\citenamefont {Davis},
  \citenamefont {Miura}, \citenamefont {Sugimoto},\ and\ \citenamefont
  {Hirao}}]{Miura1996}%
  \BibitemOpen
  \bibfield  {author} {\bibinfo {author} {\bibfnamefont {K.~M.}\ \bibnamefont
  {Davis}}, \bibinfo {author} {\bibfnamefont {K.}~\bibnamefont {Miura}},
  \bibinfo {author} {\bibfnamefont {N.}~\bibnamefont {Sugimoto}},\ and\
  \bibinfo {author} {\bibfnamefont {K.}~\bibnamefont {Hirao}},\ }\bibfield
  {title} {\bibinfo {title} {Writing waveguides in glass with a femtosecond
  laser},\ }\href@noop {} {\bibfield  {journal} {\bibinfo  {journal} {Optics
  letters}\ }\textbf {\bibinfo {volume} {21}},\ \bibinfo {pages} {1729}
  (\bibinfo {year} {1996})}\BibitemShut {NoStop}%
\bibitem [{\citenamefont {Szameit}\ and\ \citenamefont
  {Nolte}(2010)}]{Szameit2010}%
  \BibitemOpen
  \bibfield  {author} {\bibinfo {author} {\bibfnamefont {A.}~\bibnamefont
  {Szameit}}\ and\ \bibinfo {author} {\bibfnamefont {S.}~\bibnamefont
  {Nolte}},\ }\bibfield  {title} {\bibinfo {title} {Discrete optics in
  femtosecond-laser-written photonic structures},\ }\href
  {https://doi.org/10.1088/0953-4075/43/16/163001} {\bibfield  {journal}
  {\bibinfo  {journal} {Journal of Physics B: Atomic, Molecular and Optical
  Physics}\ }\textbf {\bibinfo {volume} {43}},\ \bibinfo {pages} {163001}
  (\bibinfo {year} {2010})}\BibitemShut {NoStop}%
\bibitem [{\citenamefont {Ams}\ \emph {et~al.}(2005)\citenamefont {Ams},
  \citenamefont {Marshall}, \citenamefont {Spence},\ and\ \citenamefont
  {Withford}}]{Ams2005}%
  \BibitemOpen
  \bibfield  {author} {\bibinfo {author} {\bibfnamefont {M.}~\bibnamefont
  {Ams}}, \bibinfo {author} {\bibfnamefont {G.~D.}\ \bibnamefont {Marshall}},
  \bibinfo {author} {\bibfnamefont {D.~J.}\ \bibnamefont {Spence}},\ and\
  \bibinfo {author} {\bibfnamefont {M.~J.}\ \bibnamefont {Withford}},\
  }\bibfield  {title} {\bibinfo {title} {Slit beam shaping method for
  femtosecond laser direct-write fabrication of symmetric waveguides in bulk
  glasses},\ }\href {https://doi.org/10.1364/OPEX.13.005676} {\bibfield
  {journal} {\bibinfo  {journal} {Optics Express}\ }\textbf {\bibinfo {volume}
  {13}},\ \bibinfo {pages} {5676} (\bibinfo {year} {2005})}\BibitemShut
  {NoStop}%
\bibitem [{\citenamefont {Mukherjee}\ and\ \citenamefont
  {Rechtsman}(2020)}]{Mukherjee2020a}%
  \BibitemOpen
  \bibfield  {author} {\bibinfo {author} {\bibfnamefont {S.}~\bibnamefont
  {Mukherjee}}\ and\ \bibinfo {author} {\bibfnamefont {M.~C.}\ \bibnamefont
  {Rechtsman}},\ }\bibfield  {title} {\bibinfo {title} {Observation of floquet
  solitons in a topological bandgap},\ }\href
  {https://doi.org/10.1126/science.aba8725} {\bibfield  {journal} {\bibinfo
  {journal} {Science}\ }\textbf {\bibinfo {volume} {368}},\ \bibinfo {pages}
  {856} (\bibinfo {year} {2020})}\BibitemShut {NoStop}%
\end{thebibliography}%

\end{document}


\preprint{APS/123-QED}

\title{Supplemental Material: Quantized Fractional Thouless Pumping with Solitons}

\author{Marius J\"urgensen}
 \email{marius@psu.edu}
\author{Sebabrata Mukherjee}
\author{Christina J\"org}
\author{Mikael C. Rechtsman}%
 \email{mcrworld@psu.edu}
\affiliation{%
Department of Physics, The Pennsylvania State University, University Park, Pennsylvania 16802, USA}%

\date{\today}

\maketitle


\renewcommand{\thepage}{S\arabic{page}}
\renewcommand{\thesection}{S\arabic{section}}
\renewcommand{\thetable}{S\arabic{table}}
\renewcommand{\thefigure}{S\arabic{figure}}
\renewcommand{\theequation}{S\arabic{equation}}

%

\section{Waveguide Fabrication}
The waveguides are fabricated using femto-second laser writing \cite{Miura1996,Szameit2010}. We create a permanent refractive index change by focusing high-power pulses (with approximately 400\,nJ pulse energy) into a borosilicate glass sample (Corning Eagle XG). The sample is mounted on a high-precision x-y-z translation stage (Aerotech) and translated once through the focal point with a speed of 8mm/s. The 260\,fs laser pulses are emitted by a Yb-doped fiber laser system (Menlo BlueCut) using a repetition rate of 500\,kHz at a central wavelength of 1030\,nm. The mode field diameters of the laser mode are 4.6\,mm and 5.8\,mm in horizontal and vertical direction. To shape the refractive index profile of the waveguides, we use slit-beam shaping \cite{Ams2005} with a slit width of 1.8\,mm. The pulses are focused into the sample at a depth of about 100\,$\mu$m from the surface using a lens with NA=0.4 (Thorlabs A110TM-B).

\section{Cutback method}
In order to have access to the wavefunction at different $z$-positions, we cut the sample. We make ten cuts in total, each time cutting away about 7.1\,mm, which is 1/10 of the maximum propagation distance. After each cut we polish the output facet. The uncertainty in the cut position is 0.2\,mm. To cut the sample, we define a break line by transversely moving the sample through the laser focus with identical parameters as for waveguide fabrication except higher pulse energy of about 800\,nJ. We repeat the transverse movement in 10\,$\mu$m height steps. The sample is then mechanically broken along this line and subsequently polished using a commercially available polishing device (Krelltech FLex Waveguide Polisher). During the polishing of the third cut, a crack appeared in the sample close to the output facet. The crack was cut away with the forth cut. Due to the crack, for measurements taken after the third cut, we observed about 10\% lower transmission than expected for propagation without the crack. As the additional loss occurred only a short distance away from the output facet, we do not expect significant changes in the output pattern due to the crack. We included the measured data in Fig. 3b,c without any further manipulation, but excluded it from the fit of the propagation losses.

\section{Linear Waveguide Characterization}

We characterize the linear properties of the waveguides using directional couplers. The couplers consist of two straight waveguides separated by a distance $s$ from one another. We vary the length of the second waveguide and measure the transferred intensity at the output facet as a function of the coupling length. We extract the coupling constant for each separation by fitting the analytic transfer function including an on-site detuning. The resulting values for the coupling constant and the on-site detuning are shown in Fig. 2a and b, respectively. Due to the exponential mode confinement, the coupling function is expected to decay exponentially with the separation. An exponential fit of the coupling function results in: $J(s)/\text{mm}^{-1}= 6.672 \exp(-234.8 s/\text{mm})$. We use this coupling function for the simulations shown in Figs. 1-3 and S5.

\begin{figure*}[h!]
\includegraphics{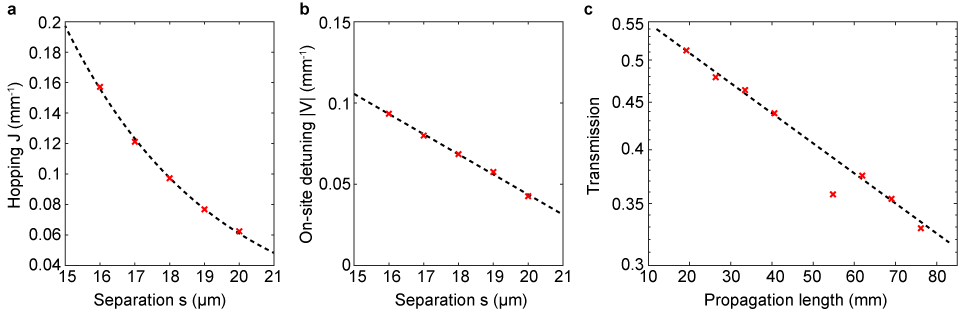}
\caption{\label{figS1} Linear waveguide characterization. \textbf{a.} Hopping constant $J$ as a function of separation $s$ between waveguides. Each $J$ is calculated via fitting 15 couplers with varying coupling length. Dashed line shows the exponential fit for the coupling function. \textbf{b.} On-site detuning in propagation constant for two waveguides written next to each other as a function of separation $s$. \textbf{c.} Transmission measurements of a straight waveguide including in- and out-coupling losses. Y-axis is plotted in log scale. The transmission measured for a propagation of $\approx$55\,mm is excluded from the exponential fit (dashed line) due to a crack in the sample creating additional losses. The resulting insertion loss is (0.33$\pm$0.02)dB/cm.}  
\end{figure*}

The on-site refractive index difference between neighboring waveguides (thus detuning between their propagation constants) is due to the writing procedure: the fabrication of the first waveguide affects the environment and therefore the fabrication of the second waveguide. As our experiments are performed in the bulk of the system where each waveguide is fabricated next to another, with separations varying only between 16.25\,$\mu$m and 18.25\,$\mu$m, the on-site detuning is small and we neglect it in the simulations. We point out that including this small on-site detuning (fitted to be $V(s)/\text{mm}^{-1}=0.291-0.0124 s/ \mu\text{m} $) does not close the band gaps of the off-diagonal AAH-model shown in Fig. 1c. Thus, it does neither change the topological properties nor affect our experimental results of fractional pumping. 

While cutting the sample to map out the propagation of the soliton in the Thouless pump model, we additionally measure the output power of straight waveguides for an input power of 1\,mW. The measured transmission for one waveguide is shown in Fig. S1c for propagation lengths longer than 19\,mm. No values were taken for a propagation distance of $\approx$40\,mm as the sample did not break uniformly at the position where the straight waveguides were located. To fit the propagation losses we exclude the measured values for a propagation distance of $\approx$55\,mm, as this value was taken with a small crack in the sample (see also Sec. 2). The resulting exponential fits for seven straight waveguides result in a propagation loss of $(0.33\pm0.02)$ dB/cm, where the error is given as one standard deviation of the individual fits.

\section{Nonlinear Waveguide Characterization \& Measurements}

\begin{figure*}[h!]
\includegraphics{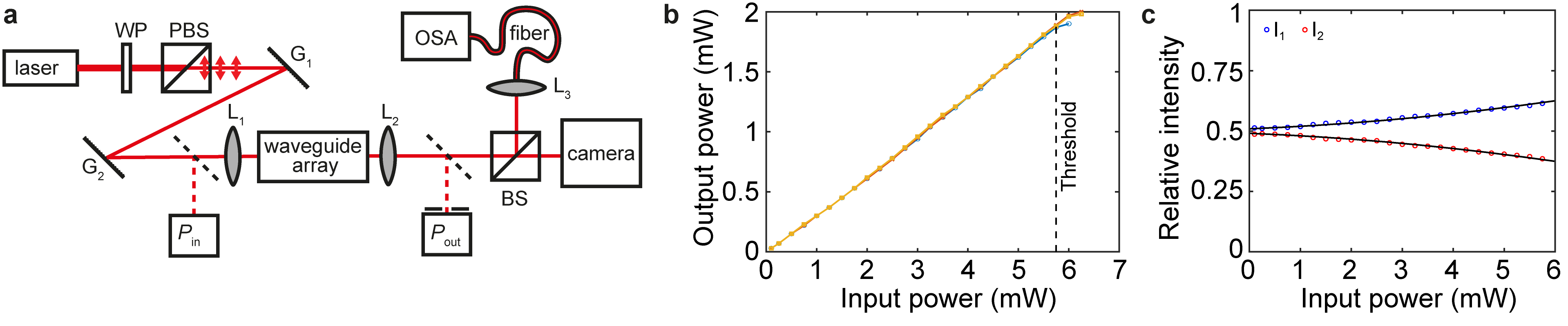}
\caption{\label{figS2} Nonlinear waveguide characterization. \textbf{a.} Setup for nonlinear characterization of the waveguides including a pair of gratings (G1 and G2) to stretch and down-chirp the excitation pulse. The output facet is imaged onto a camera and simultaneously focused into a fiber-coupled optical spectrum analyzer (OSA). \textbf{b.} Measured output power for three straight waveguides as a function of the input power showing a linear dependence and hence no nonlinear absorption for input powers less than 6\,mW. The threshold indicates the maximum power used in the experiments to prevent damage to the sample. \textbf{c.} Relative intensity in the two waveguides ($I_1$ and $I_2$) of a directional coupler as a function of power. Black lines are a least square fit, resulting in $g$=0.068/mm per mW input power.}
\end{figure*}

We use the same setup for all nonlinear measurements. A schematic (not including neutral density filters, additional mirrors and stages) is shown in Fig. S2a. A commercial laser system (Menlo BlueCut) emits 270\,fs pulses at a repetition rate of 5\,kHz (tunable), that are adjusted in power using a combination of a half-wave plate (WP) and a polarization beam splitter (PBS). A pair of gratings (G1 and G2; Thorlabs GR25-0610) down-chirps and temporally stretches the horizontally polarized pulses to 2\,ps (see also Ref. \cite{Mukherjee2020a}). The pulses are then focused into the waveguides using a lens (L1; Thorlabs AC127-030-B-ML). Lens L2 (Thorlabs AC254-040-B-ML) images the output facet of the glass sample onto a CMOS camera (Thorlabs CS165MU). Simultaneously we focus the output into a fiber (Thorlabs P1-980A-FC-1) using lens L3 (Thorlabs AC064-015-B-ML) to measure the spectrum with an optical spectrum analyzer (OSA; Anritsu MS9740 A). We use flip-mirrors to measure the time-averaged input (output) power before L1 (after L2) with a photodiode power sensor (Thorlabs S120C)

We verify the absence of nonlinear losses (for example via multi-photon absorption) by analyzing the input-output power dependence. Fig. S2b depicts the measurement for three straight waveguides. Our measurements show a linear dependence until reaching an input power of 6.0\,mW, at which the waveguides are damaged. Hence, we only use input powers $<$6.0\,mW where nonlinear losses do not play a role. 

We calibrate $g$ (see Eq. (1)) by fitting the output intensities of a directional coupler with known separation $s$=17\,$\mu$m and coupling length $l$=6\,mm. We include an on-site detuning of $V$=0.080 (see Fig. S1b) and use a hopping constant of $J$=0.131/mm. Fig. S2c shows the experimentally measured intensities, $I_1$ and $I_2$, in waveguide 1 and 2, for an excitation of waveguide 1. The corresponding output fit, using least-square fitting and including propagation losses, results in $g$=0.068/mm per mW input power, and is plotted in Fig. S2c (black).

\begin{figure*}[h!]
\includegraphics{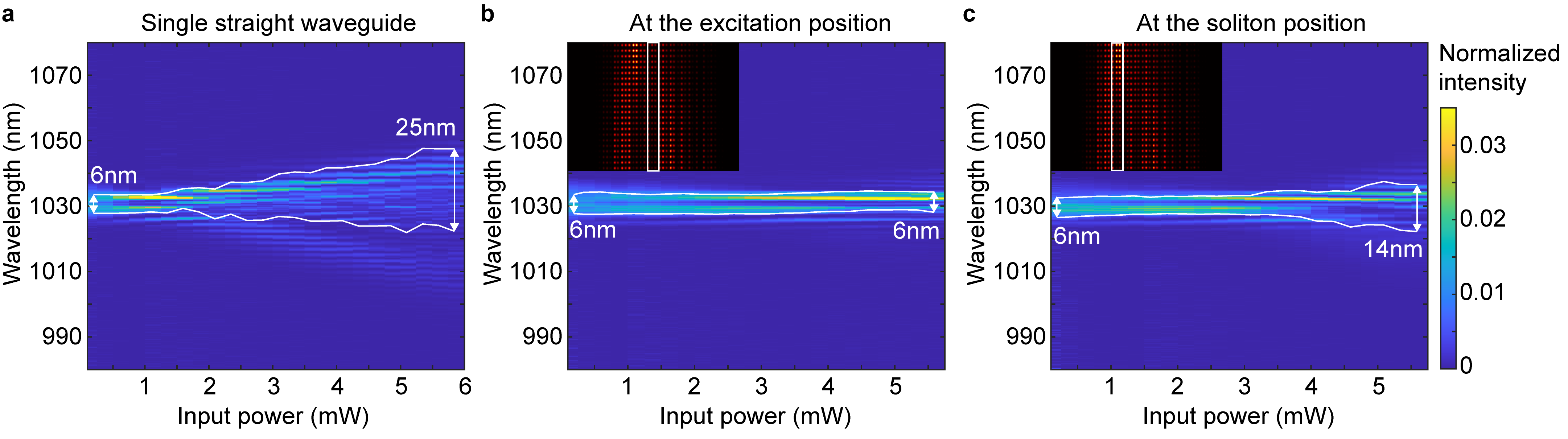}
\caption{\label{figS3} Spectral analysis. \textbf{a.} Measured normalized spectrum after propagation through a 76.15\,mm long, straight waveguide in the uncut sample for increasing input power. The white lines mark the width in which 76\% of the intensity is found (corresponding to the full-width-half-maximum of a Gaussian). \textbf{b,c.} Similar to \textbf{a} but spatially resolved and for an array showing fractional pumping. The insets show the waveguide modes at the output facet for increasing input power (from bottom to top). The white rectangle marks the spatial 1/$e^2$ width for which the spectrum is measured.} 
\end{figure*}

To accurately describe our experiment via Eq. (1), it is important that the generation of new wavelengths via self-phase modulation is minimal as the evanescent coupling is wavelength dependent. We expect maximum self-phase modulation for focusing the maximum input power into a single waveguide. Fig. S3a shows that the spectrum broadens from 6\,nm at low power to about 25\,nm at maximum input power, where this width is defined as the width in which 76\% of the intensity (equivalent to the full-width-half-maximum of a Gaussian) is found. More relevant for our experiment is the generation of new wavelengths by the fractionally pumped soliton. We measure the spatially resolved spectrum at the excitation position (Fig. S3b) and at the soliton position (Fig. S3c). Clearly, the self-phase modulation and the spectral broadening is significantly reduced to a maximum width of 14\,nm, as the fractionally pumped soliton is localized on multiple sites and therefore the power per waveguide is significantly lower. We furthermore point out that the wavelengths with the largest deviation from the central wavelength of 1030\,nm are generated close to the output facet and therefore do not significantly change the output pattern.

\section{Triple Coupler}
As it is experimentally more convenient to use single-site excitations, we use the first 5\,mm of the sample to transform a single-site excitation into an effective two-site excitation to efficiently excite the soliton. We achieve this by writing an auxiliary 5\,mm long waveguide on top of two waveguides that are part of the Thouless pump model (see Fig. S4a and also Fig. 1d). Fig. S4b shows the measured intensities in the three waveguides after 5\,mm for increasing input power. Due to the large coupling constant between the waveguides, the effect of a nonlinearly-induced on-site index change does not significantly change the output ratios. The slight imbalance between the intensities in the left and right waveguides are due to an on-site detuning stemming from the fabrication procedure. As the modulation in our system is not perfectly adiabatic, it turns out that this imbalance is favorable to excite the fractionally pumped soliton more effectively (with less background radiation) compared to a strict 50/50 excitation.

\begin{figure*}[h!]
\includegraphics{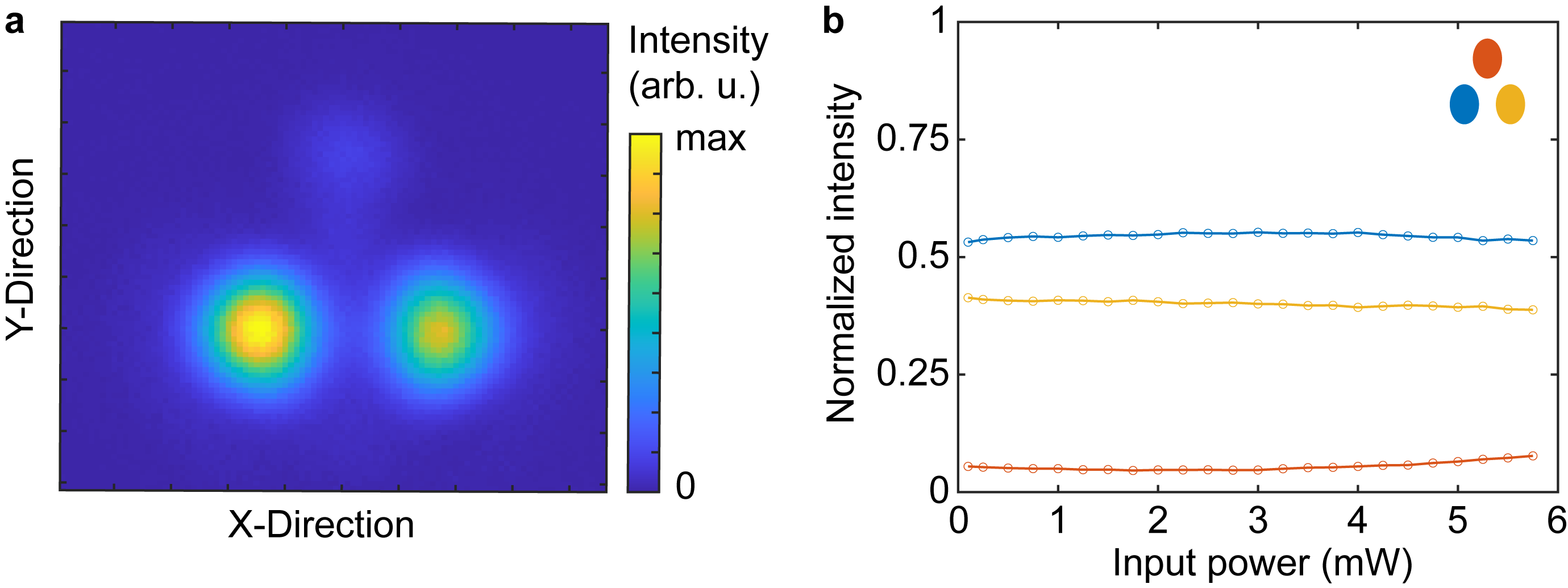}
\caption{\label{figS5} Triple coupler. \textbf{a.} Intensity distribution at the output facet for a coupler consisting of three waveguides for low input power. The coupler extends over a length of 5\,mm and the top waveguide is excited. \textbf{b.} Normalized output intensities of a triple coupler as a function of the input power. The inset defines the color-coding. Due to the high coupling constant between the waveguides over a short propagation distance of 5\,mm, the effects of nonlinearity are negligible.}  
\end{figure*}

\section{Additional information on the nonlinear transition between integer and fractionally quantized pumping}

In this section we provide more information on the transition from an integer pumped soliton to a fractionally pumped soliton for increasing nonlinearity. Figs. S5a-e show the center of mass positions of the Wannier functions of the first (red) and second (blue) band, together with the position of the relevant stable instantaneous solitons (black). Figs. S5a-e are similar to Fig. 2c, but show fewer sites and one half of a period. The behavior in the parts of the pump cycle that are not shown is identical to the parts shown due the symmetric hopping modulation and translation symmetry. For small nonlinearity ($gP/J^{\text{max}}$=0.55), i.e., in Fig. S5a, the soliton follows the path of the Wannier function of the first band, which is the band from which it bifurcates. With increasing power the solitons undergo nonlinear pitchfork bifurcations (see Fig. S5f), which for this model occur at the crossing points of the  Wannier centers of the two lowest bands. The pitchfork bifurcations split the originally contiguous path of the soliton (Figs. S5c,d) such that no quantized pumping occurs. For strong nonlinearity (Fig. S5e) a new contiguous propagation path forms, which pumps the soliton by a fractional amount per cycle. The transition from integer to fractionally quantized pumping is therefore a nonlinear transition, consistent with the pattern created by the Wannier functions of the two lowest bands.

\begin{figure*}[h!]
\includegraphics{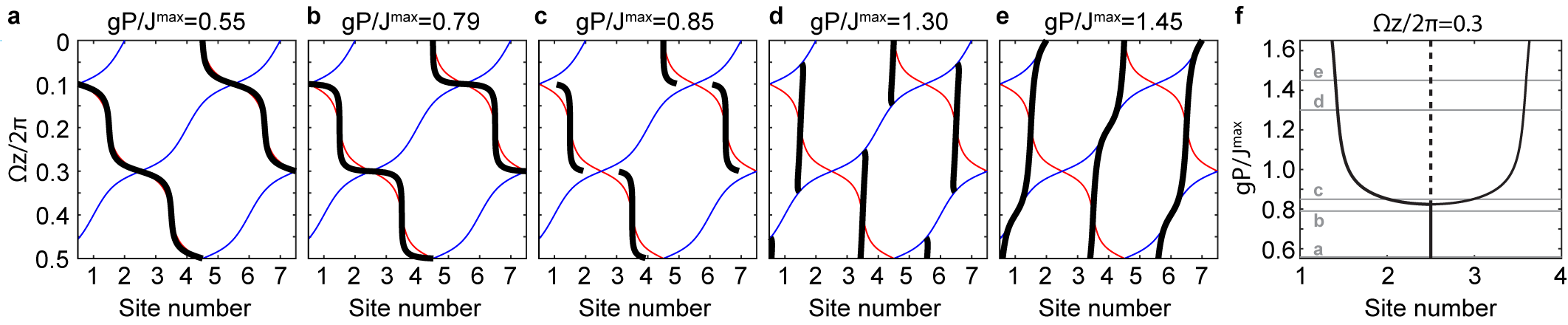}
\caption{\label{figS7} Nonlinear transition between integer and fractionally quantized pumping. \textbf{a-e.} Similar to Fig. 2c, but showing fewer sites and one half of a period. Red (blue) lines are the center of mass positions of the Wannier functions of the first (second) band. Black lines indicate the center of mass positions of stable instantaneous solitons, shown for increasing nonlinearity $gP/J^{\text{max}}$ from \textbf{a-e}. \textbf{f.} Pitchfork bifurcation diagram for $\Omega z/2\pi$=0.3 showing the stable solitons from \textbf{a-e} with solid lines, and one unstable soliton (not shown in \textbf{a-e}) with dashed lines. Horizontal gray lines indicate the strength of nonlinearity for which the $z$-evolution is shown in \textbf{a-e}.}
\end{figure*}

\section{Additional information on numerical simulation of multiple fractional plateaus within one model}

In this section we provide more details how the data for Fig. 4 in the main text is calculated. We use an off-diagonal AAH-model with 13 sites per unit cell. The on-site detuning is set to zero and the nearest-neighbor hoppings between site $n$ and $n$+1 are $J_n= K + \kappa \cos(\Omega z + \frac{10\pi}{13} n + \frac{2\pi}{13})$ with $K=1/a$ and $\kappa=0.95/a$, where $a$ is an arbitrary length. This model has 13 bands with the following Chern numbers C=\{-8,5,5,-8,5,5,-8,5,5,-8,5,5,-8\}, ordered from bottom band to top band. Due to weaker confinement, the soliton radiation is larger for low power for fixed $\Omega$. We therefore calculate the data shown in Fig. 4 differently for $gP/J^{\text{max}}<$0.06 and $gP/J^{\text{max}}>$0.06. 

For $gP/J^{\text{max}}>$0.06 the soliton is sufficiently well confined to simulate adiabatic propagation over ten periods in a lattice with 20 unit cells and a periodic length $L$=$5\cdot10^3 a$. Ten periods are a suitable choice, as they are the least common multiple of 2 and 5, such that we observe integer quantized displacement for the -3/2 and -1/5 fractional plateaus. Due to the weak soliton confinement at low power ($gP/J^{\text{max}}<$0.06), and hence the long propagation distances necessary for sufficient adiabaticity, we take advantage of the symmetries of the system and propagate those solitons only for 1/13 of a period ($L$=$5\cdot10^6 a$) in a lattice with ten unit cells. At this point, the Hamiltonian and the soliton are both identical to those at $z=0$, apart from a spatial translation, and the motion of the soliton repeats itself. Therefore, the propagation of 1/13 of the period suffices to calculate the displacement after one full period. Furthermore, we numerically check via calculations of the instantaneous soliton that the observed propagation of the first 1/13 of the period repeats for the remaining parts of the period.

\section{Description of Supplementary Animation 1}
In Supplementary Animation 1, we compare the measured propagation of the fractionally pumped soliton with numerical tight-binding simulations. The animation complements Fig. 3 and shows the integrated intensity per waveguide for two periods, measured by cutting the sample and observing the intensity distribution at the output facet for increasing input power.


\bibliography{bibliography}